\documentclass[12pt]{article}

\usepackage{amsfonts}
\usepackage{bbm}
\usepackage{cite}

\newcommand{\bmat}{\left(\begin{array}}
\newcommand{\emat}{\end{array}\right)}

\def\gtrsim{\mathrel{\raise.3ex\hbox{$>$\kern-.75em\lower1ex\hbox{$\sim$}}}}
\def\NPB#1#2#3{Nucl. Phys. B{#1} (#2) #3}

\def\ds{\displaystyle}

\def\a{\alpha}

\def\b{\beta}
\def\g{\gamma}
\def\d{\delta}

\def\om{\omega}
\def\Om{\Omega}

\def\-{\hphantom{-}}
\def\ov{\overline}
\def\s2{\frac{1}{\sqrt2}}

\def\wt{\widetilde}
\def\oh{\frac{1}{2}}
\def\beq{\begin{equation}}
\def\eeq{\end{equation}}
\def\beqa{\begin{eqnarray}}
\def\eeqa{\end{eqnarray}}
\def\D{{\rm D}}
\def\atan{{{\rm tg}^{-1} \,}}
\def\im{{\rm Im \,}}
\def\re{{\rm Re \,}}

\def\T{{\rm T}}
\def\Z{{\mathbb Z}}

\def\ca{{\cal A}}

\def\eps{\epsilon}
\def\A{\Arrowvert}

\def\cg{{\cal G}}
\def\eps{{\epsilon }}
\def\cw{{\cal W}} 
\def\cb{{\cal B}}

\def\cn{{\mathcal N}}

\def\ch{{\cal H}}
\def\cf{{\cal F}}

\def\mg{m_{3/2}}
\def\mg2{m^2_{3/2}}

\def\deq#1{\mbox{$D$=#1}}
\def\neq#1{\mbox{$\cn$=#1}}
\def\Dsl{\,\raise.15ex\hbox{/}\mkern-13.5mu D} 

\newcommand{\mathsmaller}[1]{\mbox{\footnotesize$#1$}}
\begin{document}
\pagestyle{plain}

\makeatletter
\@addtoreset{equation}{section}
\makeatother
\renewcommand{\theequation}{\thesection.\arabic{equation}}
\pagestyle{empty}
\rightline{ IFT-UAM/CSIC-05-28}
\begin{center}
\LARGE{Fluxes, moduli fixing and MSSM-like vacua in a simple IIA orientifold 
\\[10mm]}
\large{P.G. C\'amara, 
A. Font\footnote{On leave from Departamento de F\'{\i}sica, Facultad de Ciencias,
Universidad Central de Venezuela, A.P. 20513, Caracas 1020-A, Venezuela.}
and L.E. Ib\'a\~nez \\[6mm]}
\small{
Departamento de F\'{\i}sica Te\'orica C-XI
and Instituto de F\'{\i}sica Te\'orica  C-XVI,\\[-0.3em]
Universidad Aut\'onoma de Madrid,
Cantoblanco, 28049 Madrid, Spain 
\\[1mm]} 
\small{\bf Abstract} \\[1mm]
\end{center}
{\small
We study the effects of adding RR, NS and metric fluxes on a 
$\T^6/(\Omega (-1)^{F_L} I_3)$ Type  IIA orientifold. 
By using the effective flux-induced superpotential we
obtain Minkowski or AdS vacua with broken or unbroken supersymmetry.
In the Minkowski case some combinations of real moduli remain undetermined,
whereas all can be stabilized in the AdS solutions.
Many flux parameters are available which are unconstrained by
RR tadpole cancellation conditions allowing to locate the minima 
at large volume and small dilaton.
We also find that in AdS supersymmetric vacua with metric fluxes,
the overall flux contribution to RR tadpoles can vanish or have 
opposite sign to that of D6-branes,
allowing for new model-building possibilities.
In particular, we construct  the first  
$\neq1$ supersymmetric intersecting D6-brane 
models with MSSM-like spectrum and with all closed string moduli
stabilized.  Some axion-like fields remain undetermined but they
are precisely required  to give St\"uckelberg masses to
(potentially anomalous) 
$U(1)$ brane fields. We show that the cancellation of the Freed-Witten anomaly 
guarantees that the axions with flux-induced masses are orthogonal to
those giving masses to the $U(1)$'s.
Cancellation of such anomalies also guarantees that
the  D6-branes in our $\neq1$ supersymmetric AdS vacua
are calibrated so that they are forced to preserve one unbroken
supersymmetry.}


\newpage
\setcounter{page}{1}
\pagestyle{plain}
\renewcommand{\thefootnote}{\arabic{footnote}}
\setcounter{footnote}{0}

\section{Introduction}
\label{sec:intro}

One of the most pressing problems in string theory is the issue of moduli
stabilization.
Lately, important progress has been accomplished by taking into
account the freedom of switching on (quantized) RR and NS fluxes in the 
compact closed string background. This road has been particularly
explored in the context of type IIB theory, where RR/NS
fluxes create a superpotential \cite{gvw} that
depends on the complex structure fields
and the axi-dilaton and allows to fix these fields dynamically 
\cite{gkp, kst, fp, tt, blt, cu, Berg, iibcy, lrs1, ms, Cveticflux, font}.
In order to further determine the K\"ahler moduli, 
non-perturbative effects have also been put to work \cite{kklt, cynonpert}. 
For other proposals for fixing K\"ahler moduli, see \cite{otherkahler}.
In simple IIB toroidal orientifolds \cite{kst, fp, tt, blt, cu, lrs1, ms, Cveticflux, font} 
in general the moduli are fixed in
regions in which the compact volume is of order the string scale and/or the dilaton is
of order one, so that the validity of an effective 4-dimensional supergravity 
action is open to question. One of the important reasons why this is
the case is that the values of fluxes are strongly constrained
by RR tadpole cancellation conditions.
The situation is ameliorated in compactifications on IIB Calabi-Yau 
orientifolds since the flux contribution to
tadpoles is typically large and one can generate vacua 
with all moduli and the dilaton in regions of parameter space where
the effective supergravity approximation may be trusted \cite{iibcy}.

In comparison, less effort has been devoted to the similar moduli-fixing 
problem in the case of type IIA compactifications. The fact that in type IIA 
there are fluxes with both even and odd rank suggests that 
both complex structure and K\"ahler moduli fields may be determined
simultaneously without resorting to non-perturbative effects.
This has been anticipated by several authors \cite{dkpz, kk, gl, vz, DeWolfe}. 
Indeed, in  the type IIA case, RR/NS backgrounds give rise to superpotentials 
depending both on K\"ahler and complex structure moduli, 
but with no terms mixing both kinds of moduli. Furthermore, in simple toroidal
settings one can also include metric fluxes
and generate superpotential terms coupling  both kinds of moduli \cite{dkpz, vz}.
The so-called metric fluxes can arise partially from T-duality of NS fluxes
\cite{glmw, kstt, Schulz}. More generally, turning on constant metric fluxes
corresponds to Scherk-Schwarz reductions \cite{ss} that can
be understood as compactifications on twisted tori 
\cite{km, DallAgata, Andrianopoli, Reid}. 

The purpose of this paper is twofold. We will first present a detailed study
of minima of the  moduli potential induced by RR, NS and  metric fluxes in the
simple $\T^6/(\Omega (-1)^{F_L} I_3)$ type  IIA orientifold.
We concentrate on the potential for the dilaton and the diagonal K\"ahler and 
complex structure moduli, which may be also viewed as the only untwisted moduli of a
related $\Z_2\times \Z_2$ orientifold. We argue though that 
the results found for $\T^6/(\Omega (-1)^{F_L} I_3)$ ignoring off-diagonal
moduli still constitute  extrema of the potentials which are stable in 
relevant cases. We find four classes of (non-singular) vacua
which correspond to \neq1 supersymmetric models in Minkowski space, no-scale,
AdS with $\neq1$ supersymmetry and non-supersymmetric AdS models. In the Minkowski cases 
only a few of the moduli may be determined. On the other hand, the 
AdS vacua  look particularly interesting since all moduli are stabilized 
(except for a combination of axion-like fields, we come back  to this point below). 

The structure of vacua in both Minkowski and AdS space depends very much on the 
existence or not of metric fluxes which lead to some remarkable new 
features. In particular, in \neq1 supersymmetric AdS vacua {\it without}
metric fluxes,  NS and RR fluxes always contribute to RR tadpoles 
like D6-branes do. The RR tadpole cancellation conditions 
restrict some of the flux parameters but some others 
(particularly the RR 4-form and 2-form fluxes) remain unconstrained.
Due to this fact, one can easily find minima with
all closed string moduli stabilized in regions with large volume
and small dilaton, so that the effective supergravity action
should be a good approximation. The (negative) cosmological constant
may be made arbitrarily small for sufficiently large fluxes.
\neq1 supersymmetric AdS minima in generic IIA orientifolds  
with NS and RR fluxes were recently analyzed in \cite{DeWolfe}. 
In this case we obtain analogous results. 
We also find examples of AdS vacua with broken supersymmetry and all moduli 
stabilized.

In the \neq1 AdS vacua with RR/NS backgrounds {\it and} metric fluxes 
turned on a particular new property appears. The flux contribution 
to the RR tadpoles may be positive, negative or zero. This is due
to the fact that the RR-charge $Q_{RR}$ has the schematic structure 
 $Q_{RR}\simeq (m\ov{H}_3+\omega \ov{F}_2)$, where $m$ is the $0$-form
of massive IIA supergravity and $\omega$ represents the metric flux parameters.
The signs of the different fluxes are not arbitrary
since they are correlated to the signs of the real parts of the moduli
fixed at the minimum.  
The fact that we can add fluxes determining  moduli but
not contributing to RR tadpoles is important since this means that
we have a rigid `corset', namely the concrete AdS \neq1 background, 
which can be added to any  RR tadpole-free configuration of D6-branes to
stabilize  all moduli. On the other hand, considering fluxes contributing
like O6-planes to RR tadpoles 
is interesting since we can dispose of orientifold planes in certain
cases. We find, at least for the massive $m\not=0$ case, that
the AdS supersymmetric minima may be made to reside at points with large 
compact volume and small dilaton so that corrections are under control.

The observation that one can have
string backgrounds leading to vanishing or negative RR charges in AdS is not
new, see e.g.\cite{Acharya, cu2}, and is also related to the fact that
in the presence of metric fluxes one is really dealing with non-Calabi-Yau
manifolds with peculiar topology. In our case we have a twisted torus
with a half-flat structure. Then, a D6-brane 
wrapping a certain 3-cycle in the original torus with
RR charges cancelled by some background including metric fluxes
may be alternatively understood as a homologically trivial brane in the
twisted torus which is however stable because it wraps a
generalized calibrated 3-cycle \cite{cu2, juan}.

The second main topic in this paper is the inclusion of D6-branes
in models with fluxes and the construction of some semi-realistic
examples. It turns out that adding branes gives rise to some
new interesting features beyond the obvious one of their contribution
to RR tadpoles. Stacks of D6-branes wrapping 3-cycles contain in general  
$U(1)$ fields which couple to RR fields, the imaginary parts of the complex 
structure moduli and the axi-dilaton.
In particular, some of the $U(1)$'s get St\"uckelberg masses by combining
with these RR fields. Now, if some of such RR fields get masses from
fluxes some inconsistency is expected, in particular, the flux 
induced superpotential would violate gauge invariance. Therefore, we need that
the linear combinations of RR fields combining with $U(1)$'s and those getting masses
from fluxes should be orthogonal. We find that this is guaranteed as long
as the Freed-Witten  
anomaly  \cite{fw,mms} induced on the world-volume of D6-branes by
the fluxes cancels. This turns out to be an important constraint on 
the Minkowski minima. In the case of AdS \neq1 supersymmetric minima,
with the real parts of all moduli determined, 
we find the interesting result that the cancellation of the FW
anomaly  automatically forces the branes to preserve supersymmetry, i.e. to 
wrap special Lagrangian (slag)  cycles. Stating it the other way around, any 
D6-brane wrapping slag cycles will automatically be free of the FW anomaly
in this background.

It is important to see how far one can go in stabilizing  all moduli in 
models with possible phenomenological relevance.
We present examples of configurations of D6-branes wrapping 3-cycles
on the torus and intersecting at angles with chiral MSSM-like spectra and
fixed moduli. The models contain three generations of quarks and leptons and 
one Higgs set, with the gauge group of the SM extended by one or two extra 
$U(1)$'s and some extra heavy vector-like $SU(2)_L$ doublets and singlets. 
Some of the examples live in Minkowski space (either \neq1 supersymmetric or 
no-scale), in which case only a few moduli are fixed. 
On the other hand, we present the first  
semi-realistic \neq1 supersymmetric model in AdS with all closed string
moduli stabilized. This model requires the presence of
both metric and NS/RR fluxes so that a `wrong sign' contribution
to tadpoles, mimicking orientifold planes, is obtained.
The minima are stabilized in a perturbative regime and 
all physical quantities like gauge and Yukawa couplings are given in terms
of otherwise undetermined NS and RR fluxes.

The outline of the paper is as follows. In the next section we introduce 
the basic tools needed to describe the 
$\T^6/(\Omega (-1)^{F_L} I_3)$ type IIA orientifold with fluxes.
In section 3 we display the structure of the flux-induced superpotential
and discuss the contribution of different fluxes to RR charges. The
systematic analysis of the different vacua of the flux-induced potential
is presented in section 4. Examples of Minkowski and AdS vacua with/without supersymmetry  
and with/without metric fluxes are reported. In section 5 we examine the
constraints coming from the Freed-Witten anomaly and their connection
to the open string $U(1)$'s. Specific semi-realistic intersecting D6-brane models
with MSSM-like spectrum are discussed in section 6. This includes the 
AdS \neq1 supersymmetric example with all closed string moduli stabilized. 
We present some comments and conclusions in section 7. 
Some related results are collected in two appendices. In appendix A
we study the $SU(3)$ structure of the twisted torus and discuss \neq1 vacua
in terms of torsion classes.  In appendix B we present some 
non-supersymmetric D6-brane configurations with moduli stabilized  in AdS.

\section{Basic Features}

The aim of this section is to present the concepts needed to
describe the low-energy effective action of type IIA orientifolds 
in the presence of background fluxes. We first introduce the moduli fields 
and then exhibit the superpotential induced by NS and RR fluxes.
We next define the so-called metric fluxes and recall how they
can partially arise from T-duality of NS fluxes in type IIB.

\subsection{IIA orientifolds: moduli and NS/RR fluxes}
\label{sec:orientifold}

This section contains a brief review of the structure of IIA Calabi-Yau
orientifolds. The treatment follows \cite{gl} which
the reader can consult for more details. We limit to a discussion
of moduli fields, NS/RR fluxes, and flux-induced superpotentials.
Our main purpose is to apply the general results to a simple IIA
toroidal orientifold.

Compactification of type IIA strings on a Calabi-Yau 3-fold $Y$ gives
a \deq4, \neq2 theory with $h_{11}$ vector multiplets and $(1+h_{12})$
hypermultiplets \cite{cadavid}. Turning on fluxes for the NS and RR
field strengths generates a potential for the scalars in these
multiplets \cite{ps, lm, kk}.  To obtain a \neq1 theory one can
implement an orientifold projection by $\Omega_P (-1)^{F_L} \sigma$,
where $\Omega_P$ is the world-sheet parity operator, $(-1)^{F_L}$ is
the space-time fermionic number for left-movers and $\sigma$ is an
order two involution of $Y$. The action of $\sigma$ on the K\"ahler
form and the holomorphic 3-form is $\sigma(J)=-J$ and
$\sigma(\Omega)=e^{2i\theta} \Omega^*$.  We take $\theta=0$ and
$\sigma(z^i)= \bar z^i$, where $z^i$ are local complex coordinates.
This implies O6-planes whose tadpoles can be cancelled by adding
D6-branes or flux, as we will see.

The closed (1,1) forms split into $h_{11}^+$ and $h_{11}^-$, according
to whether they are even or odd under $\sigma$. There is an equal
number $(1+h_{12})$ of even and odd 3-forms. Then, the resulting
matter content from the closed string sector consists of $h_{11}^+$
vector multiplets, $h_{11}^-$ chiral multiplets corresponding to
K\"ahler moduli, and $(1+h_{12})$ chiral multiplets corresponding to
the dilaton and the complex structure moduli \cite{gl}. The scalar 
components of the K\"ahler moduli, denoted $T_A$, are defined in terms 
of the complexified K\"ahler form as 
\beq 
J_c = B + iJ = i
\sum_{A=1}^{h_{11}^-} T_A \omega_A \ ,
\label{complexj}
\eeq 
where $\omega_A$ are the $\sigma$-odd (1,1) closed forms. The
complex structure moduli, denoted $N_L$, $L=0, \cdots, h_{12}$, can be
extracted from 
\beq 
iN_L = \int_Y \Omega_c \wedge \beta_L \quad ;
\quad \Omega_c= C_3 + i \re (C\Omega) \ ,
\label{omegac}
\eeq 
where $C_3=\sum_L \xi_L \a_L$ provides the axions. Here $\a_L$
and $\b_L$ are respectively the $\sigma$-even and $\sigma$-odd
3-forms. The field $C$ is in turn specified by 
\beq 
C= e^{-\phi_4}
e^{K_{cs}/2} \quad ; \quad K_{cs}=-\log [-\frac{i}8 \int_Y \Omega
\wedge \Omega^* ] \ ,
\label{cfield}
\eeq 
where $\phi_4$ is the T-duality invariant four-dimensional
dilaton given by $e^{\phi_4}= e^{\phi_{10}}/\sqrt{{\rm vol}\, Y}$.

To be more explicit let us now consider the example of a factorized
6-torus $\otimes_{j=1}^3 \T_j^2$. The area of each sub-torus is
$(2\pi)^2 A_j$, where $A_j = R_x^j R_y^j$. The $A_j/\a^\prime$ are thus the
real part, $t_j$, of three K\"ahler moduli $T_j$. For each sub-torus we take a
square lattice, consistent with the orientifold projection. The
complex structure parameter of each $\T_j^2$ is then
$\tau_j=R_y^j/R_x^j$. It is known, see e.g. \cite{cim1}, that in
this setup the IIA \deq4 fields $S$ and $U_i$, corresponding to the
dilaton and complex structure moduli, have real parts 
\beq 
\re S \equiv s = \frac{e^{-\phi_4}}{\sqrt{\tau_1 \tau_2 \tau_3}} \quad ;
\quad \re U_i \equiv u_i = e^{-\phi_4} \sqrt{\frac{\tau_j
\tau_k}{\tau_i}} \quad ; \ i\not=j\not=k \ ,
\label{csmoduli}
\eeq 
where $e^{\phi_4}= e^{\phi}/\sqrt{t_1 t_2 t_3}$. We will
next obtain these results from the general analysis of \cite{gl}.  As
usual, the holomorphic 3-form can be written as 
\beq 
\Omega=(dx^1 + i\tau_1\, dy^1) \wedge (dx^2 + i\tau_2\, dy^2) 
\wedge (dx^3 + i\tau_3\, dy^3) \ ,
\label{holo} 
\eeq 
where $y^i=x^{i+3}$. The orientifold involution acts as
$\sigma(x^i)=x^i$ and $\sigma(y^i)=-y^i$.  The even and odd
3-forms with one leg on each sub-torus are
\beqa 
\a_0 & = & dx^1 \wedge dx^2 \wedge dx^3 \quad ; \quad 
\b_0 = dy^1 \wedge dy^2 \wedge dy^3 \ , \nonumber \\[0.2cm] 
\a_1 & = & dx^1 \wedge dy^2 \wedge dy^3 \quad ; \quad 
\b_1 = dy^1 \wedge dx^2 \wedge dx^3 \ , \label{abbasis} \\[0.2cm] 
\a_2 & = & dy^1 \wedge dx^2 \wedge dy^3 \quad ; \quad 
\b_2 = dx^1 \wedge dy^2 \wedge dx^3 \ , \nonumber \\[0.2cm] 
\a_3 & = & dy^1 \wedge dy^2 \wedge dx^3 \quad ; \quad 
\b_3 = dx^1 \wedge dx^2 \wedge dy^3 \ . \nonumber 
\eeqa 
Our normalization is such that $\int_{\T^6} \a_I \wedge \b_J = \d_{IJ}$.
Substituting in (\ref{cfield}) we find $C= \re S$. From (\ref{omegac})
we then obtain the corresponding moduli $N_0=S$ and $N_i=-U_i$.

We have only considered 3-forms with one leg on each 
$\T_j^2$ because these are the directions in which we are going
to switch on fluxes. If the orientifold has an extra  $\Z_2 \times \Z_2$ symmetry
these are in fact the only invariant forms.

The next step is to turn on background fluxes. The NS $H_3$ is odd
under the orientifold action, thus the general flux allowed is
\beq 
\ov{H}_3 = \sum_{L=0}^{h_{12}} h_L \b_L \ .
\label{hfluxes}
\eeq 
For the RR forms, $F_0$ and $F_4$ are even while $F_2$ and $F_6$
are odd under the orientifold projection. We can then have the general
expansions 
\beqa 
\ov{F}_0 = -m & ; & \ov{F}_6 = e_0 d{\rm vol}_6
\nonumber \\[0.2cm] \ov{F}_2 = \sum_{A=1}^{h_{11}^-} q_A \om_A & ; &
\ov{F}_4 = \sum_{A=1}^{h_{11}^-} e_A \wt{\om}_A \ ,
\label{ffluxes}
\eeqa 
where $\wt{\om}_A$ are the $h_{22}^+=h_{11}^-$ $\sigma$-even
(2,2) forms.  There are also quantization conditions 
\beq
\frac{\ell^3 \mu_1}{2\pi} \int_{\Pi_3} \ov{H}_3 \in \Z \quad ; \quad
\frac{\ell^p \mu_{p-2}}{2\pi} \int_{\Pi_p} \ov{F}_p \in \Z \ ,
\label{qcon}
\eeq 
for any $p$-cycle $\Pi_p$ in $Y$. Here 
$\ell=2\pi\sqrt{\a^\prime}$, and $\mu_p=1/(2\pi)^p
\a^{\prime \, (p+1)/2}$ \cite{poldbranes, boussopol}.  We normally
take the various forms, e.g. $\b_L$, to belong to an integer basis so
that in units of $2\pi/\mu_{p-2} \ell^p=1/\ell$ the various coefficients, such as
$h_L$, are integers. Actually, to avoid subtleties with exotic
orientifold planes \cite{gkp,kst} we take the coefficients to be even.
Notice that by including the factors of $\ell$ explicitly, all forms have
dimensions $({\rm length})^{-1}$. With these conventions the moduli
fields are all dimensionless.

The RR fluxes generate a superpotential for the K\"ahler moduli that
can be written as 
\beq 
W_K = 
\int_Y e^{J_c} \wedge \ov{F}_{RR} \ ,
\label{wkahler}
\eeq 
where $\ov{F}_{RR}$ represents a formal sum of the even RR
fluxes. This result can be obtained \cite{tv} applying mirror symmetry
to the type IIB superpotential \cite{gvw}. It can also be derived
performing the explicit Kaluza-Klein reduction \cite{gl} which also
allows to determine the superpotential for the complex structure
moduli due to NS flux, namely 
\beq 
W_Q = \int_Y \Omega_c \wedge \ov{H}_3
= i \sum_{L=0}^{h_{12}} h_L N_L \ ,
\label{wcs}
\eeq 
where we have used (\ref{omegac}). The K\"ahler potential for
both kinds of moduli are given by 
\beq 
K_K = -\log[\frac43 \int_Y J
\wedge J \wedge J] \quad ; \quad K_Q = - \log e^{-4\phi_4} \ .
\label{kpotgeneral}
\eeq 
There are corrections to $K_K$ due to world-sheet instantons and
to $K_Q$ due to D2 instantons \cite{gl}. We will see later that
one can locate the minima of the potential in regions with 
large volume and small dilaton in which these corrections should
be in principle under control.

It is instructive to apply these results to the $\otimes_{j=1}^3
\T_j^2$ example.  The $\b_L$ are given in (\ref{abbasis}) whereas 
\beq
\om_i = -dx^i \wedge dy^i \quad ; \quad 
\wt{\om}_i = dx^j \wedge dy^j \wedge dx^k \wedge dy^k
\quad ; \ i\not=j\not=k \ .
\label{omomt}
\eeq 
Notice that $\int_{\T^6} \om_i \wedge \wt{\om}_j = \d_{ij}$.  It
is straightforward to substitute the flux expansions in
(\ref{wkahler}) and (\ref{wcs}) to obtain 
\beqa 
W_K & = & e_0 +
i\sum_{i=1}^3 e_i T_i - q_1 T_2 T_3 -q_2 T_1 T_3 -q_3 T_1 T_2 + i m
T_1 T_2 T_3 \ , \nonumber \\[0.2cm] W_Q & = & ih_0S - i\sum_{i=1}^3
h_i U_i \ .
\label{sup0}
\eeqa 
These superpotentials have been recently discussed in
\cite{dkpz, vz}.  Finally, the K\"ahler potential takes the usual
expression 
\beq 
K= - \log(S+S^*) - \sum_{i=1}^3 \log(U_i + U_i^*) -
\sum_{i=1}^3 \log(T_i + T_i^*) \ .
\label{kpot}
\eeq

\subsection{Metric fluxes and twisted tori}
\label{sec:metric}

In the next section we will see how superpotential terms mixing K\"ahler and 
complex structure moduli, including the dilaton, can be generated by switching
on so-called metric fluxes. Such backgrounds appear naturally in the context of
Scherk-Schwarz reductions \cite{ss}. In turn these can be shown (see e.g. \cite{km}) 
to be equivalent to compactification on a twisted torus defined by 
\beq
d\eta^P = -\oh \omega^P_{MN} \eta^M \wedge \eta^N \ ,
\label{ocon}
\eeq 
where $\omega^P_{MN}$ are constant coefficients, antisymmetric in the lower indices. 
The structure constants $\omega^P_{MN}$ are the metric fluxes we are
interested in. The $\eta^P$ are the tangent 1-forms and can depend linearly on the 
internal coordinates $x^M$, concretely
\beq \eta^M = N_N^{\ \ M}(x) dx^N \quad ; \quad
dx^N=N^N_{\ \ M}(x) \eta^M \ .
\label{nmat}
\eeq
One can define isometry generators as
\beq Z_M
= N^N_{\ \ M} \frac{\partial}{\partial x^N} \ .
\label{zgen}
\eeq
The metric fluxes are actually the structure constants of the Lie
algebra generated by the $Z_M$, i.e.  
\beq [Z_M, Z_N] = \omega^P_{MN}
Z_P \ .
\label{zal}
\eeq
Either from the Jacobi identity of the algebra or from the
Bianchi identity of (\ref{ocon}) one finds that the metric fluxes must
satisfy
\beq \omega^P_{[MN} \omega^S_{R]P} = 0 \ .
\label{jac}
\eeq
It can further be shown that $\omega^P_{PN}=0$ \cite{ss}.

We can derive a helpful result for the exterior
derivative of a 2-form $X=X_{MN} \eta^M \wedge \eta^N$ using (\ref{ocon}). 
For coefficients independent of the $x^N$ we readily find
\beq
(dX)_{LMN} = \omega^P_{[LM} X_{N]}{}_P   \  .
\label{dXomega}
\eeq
Similar formulas can be obtained for higher forms.

We will focus on the case in which only metric fluxes of type
$\omega^i_{ab}$, $\omega^i_{jk}$, $\omega^a_{ib}$, $i=1,2,3$,
$a=4,5,6$, are allowed. This can be implemented by imposing a symmetry
of (\ref{ocon}) under the orientifold involution $\eta^i \to \eta^i$,
$\eta^a \to -\eta^a$. As in the case of RR and NS fluxes discussed
previously, we are only going to switch on metric fluxes along factorized
directions. These correspond to the structure constants 
which are invariant under a $\Z_2\times \Z_2$ symmetry 
whose generators
transform $(\eta^1, \eta^2, \eta^3, \eta^4, \eta^5, \eta^6)$ into
$(-\eta^1, -\eta^2, \eta^3, -\eta^4, -\eta^5, \eta^6)$ and $(\eta^1,
-\eta^2, -\eta^3, \eta^4, -\eta^5, -\eta^6)$. In the end there are
twelve metric fluxes left. To write down the relations that follow
from (\ref{jac}) we introduce the notation 
\beq 
\bmat{c} a_1 \\ a_2 \\
a_3 \emat = \bmat{c} \om^1_{56} \\ \om^2_{64} \\ \om^3_{45} \emat
\quad ; \quad \bmat{ccc} b_{11} & b_{12} & b_{13}§ \\ b_{21} & b_{22}
& b_{23} \\ b_{31} & b_{32} & b_{33} \emat = \bmat{ccc} \! \! \!
-\om^1_{23} & \, \om^4_{53} & \, \om^4_{26} \\ \, \om^5_{34} & \! \!
\! -\om^2_{31} & \, \om^5_{61} \\ \, \om^6_{42} & \, \om^6_{15} & \!
\! \! -\om^3_{12} \emat \ .
\label{abmatrix}
\eeq 
The Jacobi identities imply the twelve constraints 
\beqa 
b_{ij}
a_j + b_{jj} a_i & = & 0 \quad ; \quad i \not= j \nonumber \\[0.2cm]
b_{ik} b_{kj} + b_{kk} b_{ij} & = & 0 \quad ; \quad i \not= j \not= k
\ .
\label{jacb} 
\eeqa 
There are some obvious solutions of these constraints. For
instance, (1): $b_{ij}=0$, $\forall i,j$; (2): $a_i=0$, $b_{ij}=b_i
\delta_{ij}$; (3): $a_i=a$, $b_{ij}=b$, $i \not= j$, $b_{ii}=-b$.

It is also enlightening to see how
the twisted torus structure arises by T-dualizing a string
background including constant NS-NS 3-form flux \cite{glmw, kstt}. To
simplify the discussion, as internal space we take the flat torus
$\T^6$, i.e. we neglect the warp factors needed to have a solution of
the equations of motion.  Concretely, we start from a type IIB background 
\beqa 
ds^2 & = &
(dx^1)^2 + \cdots + (dx^6)^2 \ , \nonumber \\[0.2cm] \ch_3 & = & -
a_1\ dx^1 \wedge dx^5 \wedge dx^6 - a_2\ dx^4 \wedge dx^2 \wedge dx^6
- a_3\ dx^4 \wedge dx^5 \wedge dx^3 \ .
\label{ghdata}
\eeqa 
We want to perform T-dualities in $x^1, x^2, x^3$. For the
magnetic field we choose a gauge such that $\cb_2$ does not depend on
the dualized coordinates. We take 
\beq 
\cb_2 = - a_1\, x^6 \ dx^1
\wedge dx^5 - a_2\, x^4 \ dx^2 \wedge dx^6 - a_3\, x^5\ dx^3 \wedge
dx^4 \ .
\label{bdata}
\eeq 
Using standard results (see e.g. appendix A in \cite{kstt}) gives
the transformed metric 
\beqa 
ds^{\prime 2} & = & (dx^1 + a_1 x^6
dx^5)^2 + (dx^2 + a_2 x^4 dx^6)^2 + (dx^3 + a_3 x^5 dx^4)^2 
\nonumber \\[0.2cm] 
&+& (dx^4)^2 + (dx^5)^2 + (dx^6)^2 \ .
\label{gprime}
\eeqa 
Moreover, $\ch^{\prime}_3 = 0$. All of the NS-NS flux is traded
for metric flux. {}From the new metric we read off the following
tangent 1-forms 
\beqa 
\eta^1 & = & dx^1 + a_1 x^6 dx^5 \quad ; \quad
\eta^4=dx^4 \ , \nonumber \\[0.2cm] 
\eta^2 & = & dx^2 + a_2 x^4 dx^6
\quad ; \quad \eta^5=dx^5 \ , \label{netas} \\[0.2cm] 
\eta^3 & = &
dx^3 + a_3 x^5 dx^4 \quad ; \quad \eta^6=dx^6 \ . \nonumber 
\eeqa
Taking the exterior derivatives 
we easily identify the structure
constants: $\omega^1_{56}= a_1$, $\omega^2_{64}= a_2$, and
$\omega^3_{45}= a_3$.

Finally, an important point is that metric fluxes are also quantized.
For the $a_i$ we have just seen that they are obtained from T-duality
of NS fluxes. In general this is needed for consistency of the
twisted torus structure \cite{lu}.

\section{IIA superpotential and RR tadpoles due to general fluxes}
\label{sec:sup}

It is instructive to check how a number of terms in the IIA
superpotential (including some induced by metric fluxes) 
may be obtained  applying T-duality transformations to the known type
IIB results.  Our starting point is the type IIB orientifold
$\T^6/\Omega (-1)^{F_L} I_6$, where $I_6$ reflects the six internal
coordinates $x^M$. There are 64 O3-planes whose charge can be
cancelled by adding D3-branes and/or flux. To go to type IIA we will
implement mirror symmetry which is the same as T-duality in $x^1, x^2,
x^3$. In the type IIA picture there are then O6-planes, wrapping the
$x^i$, $i=1,2,3$, and one can add intersecting D6-branes.

We consider a factorized geometry in which $\Omega$ is given in
(\ref{holo}).  In order to generate a superpotential for the $\tau_i$
and the axion-dilaton we turn on NS $\ov{\ch}_3$ and RR $\ov{\cf}_3$ 3-form
fluxes that are conveniently expanded in the basis (\ref{abbasis}).
For $\ov{\cf}_3$ we take the most general combination 
\beq 
\ov{\cf}_3=-m \a_0
-e_0\b_0 + \sum_{i=1}^3 (e_i\a_i - q_i \b_i) \ .
\label{fflux}
\eeq 
Mirror symmetry transforms $\ov{\cf}_3$ into RR fluxes
$(\ov{F}_0,\ov{F}_2,\ov{F}_4,\ov{F}_6)$ in type IIA.  For  $\ov{\ch}_3$ we 
instead restrict to
\beq 
\ov{\ch}_3 = h_0\b_0 - \sum_{i=1}^3 a_i\a_i \ .
\label{hflux}
\eeq 
Under the three T-dualities only $h_0$ remains as NS flux,
i.e. $\ov{\ch}_3 \to \ov{H}_3=h_0 \b_0$.  As discussed before, the $a_i$ become
instead metric fluxes, in fact $\omega^1_{56}= a_1$, $\omega^2_{64}=
a_2$ and $\omega^3_{45}= a_3$. We do not turn on $\ov{\ch}_3 \sim \a_0$
because then $\cb_2$ would depend on the $x^i$. 
We do not consider 
 $\ov{\ch}_3 \sim
\b_i$  fluxes either  because they lead to more complicated metrics
\cite{kstt, nongeo, Reid}.

The type IIB superpotential induced by the fluxes is given by
\cite{gvw} 
\beq 
\cw = \int (\ov{\cf}_3 - \tau \ov{\ch}_3) \wedge \Omega \ .
\label{wbf}
\eeq 
Here $\tau=C_0 + ie^{-\phi}$, where $C_0$ is the RR 0-form.
Substituting (\ref{holo}), (\ref{fflux}) and (\ref{hflux}) we find 
\beq 
\cw = e_0 + h_0 \tau + i\sum_{i=1}^3 (e_i+a_i\tau)\tau_i -
q_1 \tau_2 \tau_3 - q_2 \tau_1 \tau_3 - q_3 \tau_1 \tau_2 +i m
\tau_1\tau_2\tau_3 \ .
\label{wb}
\eeq 
Upon mirror symmetry the $\tau_i$ go into K\"ahler moduli and
$\tau$ becomes the IIA dilaton, $\tau \to iS$. Hence, we obtain the
IIA superpotential 
\beq 
W_{ST} = e_0 + ih_0 S + \sum_{i=1}^3 (ie_i-a_i
S)T_i - q_1 T_2 T_3 - q_2 T_1 T_3 - q_3 T_1 T_2 +i m T_1 T_2 T_3 \ .
\label{wast}
\eeq 
Notice that for $a_i=0$ this coincides with (\ref{sup0}), with
$h_i=0$, that was derived following the analysis of \cite{gl}. For
$a_i \not=0$ it agrees with results of \cite{dkpz, vz}.

In type IIB the fluxes contribute to the $C_4$ tadpole with
coefficient 
\beq 
N_{flux} = \int \ov{\ch}_3 \wedge \ov{\cf}_3 = h_0 m + a_1 q_1
+ a_2 q_2 + a_3 q_3 \ ,
\label{nf3}
\eeq 
where we already substituted the fluxes at hand. Under mirror
symmetry $N_{flux}$ transforms into a $C_7$ tadpole in the direction
of the O6-planes.  
This tadpole also receives contributions from D6-branes. In general
we introduce piles of $N_a$ intersecting D6-branes wrapping the 
factorizable 3-cycle
\beq
\Pi_a=(n_a^1, m_a^1)\otimes(n_a^2, m_a^2) \otimes(n_a^3, m_a^3) \  ,
\label{wns}
\eeq
and the corresponding orientifold images wrapping $\otimes_i (n_a^i, -m_a^i)$.
Here $n_a^i$ $(m_a^i)$ are the wrapping numbers along the $x^i$ $(y^i)$
torus directions.
Including the O6-planes, that wrap $\otimes_i(1,0)$, leads to
the tadpole cancellation condition
\beq
\sum_a N_a n_a^1 n_a^2 n_a^3 + \oh(h_0 m + a_1 q_1 + a_2 q_2 + a_3 q_3) = 
16 \  .
\label{tadxxx}
\eeq
This agrees with the result of \cite{vz}.
Tadpoles due to fluxes of the NS and the RR 0-form have been considered in
\cite{Uranga, kk, DeWolfe}. To our knowledge, tadpoles due to metric fluxes
were first discussed in \cite{cu}. 

The tadpole condition can also be derived from the equation of motion
for $C_7$ in type IIA. Let $G_2=dC_1 + m B_2 + \ov{F}_2$ and
${}^*F_2=F_8=dC_7$, then the relevant piece in the action is 
\beq
\int_{M_4 \times Y}[C_7\wedge dF_2 + C_7 \wedge (m \ov{H}_3 +
d\ov{F}_2 )] + \sum_a N_a \int_{M_4 \times \Pi_a} C_7 \ .
\label{c7tad}
\eeq 
The first term arises from the kinetic energy $\int G_2 \wedge
{}^*G_2$.  The second term takes into account the coupling to
D6-branes and O6-planes. The important point to notice is that 
$d\ov{F}_2 \not=0$ due to the metric fluxes. For instance, using
(\ref{dXomega}), we obtain $(d\ov{F}_2)_{456}=(a_1 q_1 + a_2 q_2 + a_3
q_3)$, and thus recover (\ref{tadxxx}). Moreover, from other
components of $C_7$ there are further cancellation conditions 
\beqa
\sum_a N_a n_a^1 m_a^2 m_a^3 + \oh (m h_1 - q_1 b_{11} - q_2 b_{21} -
q_3 b_{31}) & = & 0 \ , \nonumber \\[0.2cm] 
\sum_a N_a m_a^1 n_a^2
m_a^3 + \oh( m h_2 - q_1 b_{12} - q_2 b_{22} - q_3 b_{32}) & = & 0 \ ,
\label{tadodh} \\[0.2cm]
\sum_a N_a m_a^1 m_a^2 n_a^3 + \oh (m h_3 - q_1 b_{13} - q_2 b_{23} -
q_3 b_{33}) & = & 0 \ . \nonumber 
\eeqa 
These also agree with the conditions found in \cite{vz}.

We have just seen that the metric fluxes $b_{ij}$ create RR tadpoles.
Recently it has been observed \cite{dkpz, vz} that they also generate
superpotential terms involving the $U_k$, as expected from the fact
that the fluxes $a_i$ produce terms involving $S$. Performing a
generalized dimensional reduction on the twisted torus it has been
shown in \cite{vz} that the superpotential for the complex structure
moduli is an extension of $W_Q$ that can be expressed as 
\beq 
W_Q= \int_Y \Omega_c \wedge (\ov{H}_3 + dJ_c) \ .
\label{wqgen}
\eeq 
Such an expression was already proposed in \cite{glmw}.
The metric fluxes appear in $dJ_c$ that is computed using
(\ref{dXomega}).  A similar modification of the superpotential occurs in
heterotic compactifications on non-K\"ahler manifolds
\cite{hetflux}. In our setup, computing $W_Q$ and combining with
$W_K$ in (\ref{sup0}) yields the full superpotential 
\beqa 
W & = & e_0
+ ih_0 S + \sum_{i=1}^3 [(ie_i - a_i S - b_{ii}U_i -\sum_{j\not= i}
b_{ij}U_j)T_i - i h_iU_i] \nonumber \\[0.2cm] & - & q_1 T_2 T_3 -q_2
T_1 T_3 -q_3 T_1 T_2 + i m T_1 T_2 T_3 \ .
\label{wa}
\eeqa 
This is the result obtained in \cite{vz}.

An obvious and interesting question is how the above superpotential
changes when we include non-diagonal moduli related to parameters $\tau_{ia}$
and $t_{ia}$ that can appear in $\Omega$ and $J$. In (\ref{wa}) only
the diagonal parameters, i.e. $\tau_i\equiv \tau_{i,i+3}$ and 
$t_i\equiv t_{i,i+3}$, are taken
into account. Furthermore, we have only switched on diagonal fluxes,
along (\ref{abbasis}) and (\ref{omomt}),
which we still continue to do. Now, from
the general expressions of $W_K$ and $W_Q$,
eqs. (\ref{wkahler}) and (\ref{wqgen}), it is easy to see that the 
diagonal fluxes $h_i$, $e_i$ and $a_i$, or $h_0$, do not excite
the off-diagonal moduli, generically denoted $T^\prime$ and
$U^\prime$. However, the fluxes $m$, $q_i$ and $b_{ij}$ can
potentially generate a superpotential for $T^\prime$ and
$U^\prime$ that is at least quadratic in these fields. The
K\"ahler potential has also quadratic corrections. The upshot
is that when we look for supersymmetric minima there is always
a solution $T^\prime=U^\prime=0$ and the diagonal
moduli fixed as when $T^\prime$ and $U^\prime$ are not included. 
We know that at the point $T^\prime=U^\prime=0$ there is a global
$\Z_2 \times \Z_2$ symmetry and furthermore,
supersymmetry guarantees that this minimum is stable. 
In the following we will then disregard the off-diagonal moduli. 

For future purposes we define 
\beqa 
\wt{T}_I & = &(i, T_1, T_2, T_3)
\quad ; \quad A_{IJ} = \bmat{cccc} \!\!\! -h_0 & h_1 & h_2 & h_3 \\ \,
a_1 & b_{11} & b_{12} & b_{13} \\ \, a_2 & b_{21} & b_{22} & b_{23} \\
\, a_3 & b_{31} & b_{32} & b_{33} \emat \label{amatrix} \\[-0.4cm]
\wt{U}_I & = &(S, U_1, U_2, U_3) \ .  \nonumber 
\eeqa 
The $\wt{U}_I$ dependent superpotential, due to NS and metric fluxes, takes the
simple form 
\beq 
W_Q= -\sum_{I,J=0}^3 A_{IJ} \wt{T}_I \wt{U}_J \ .
\label{tildew}
\eeq 
The flux contribution to $C_7$ tadpoles can also be written in
terms of the matrix $A$.

Recall that the metric fluxes are constrained by the Jacobi identities
(\ref{jacb}).  For instance, there is a solution $b_{ji}=b_i$,
$b_{ii}=-b_i$, $a_i=a$.  Further choosing RR fluxes $q_i=-c_2$ and
$e_i=c_1$, allows a configuration with $T_1=T_2=T_3=T$. Then the
superpotential reduces to 
\beq 
W=e_0 + 3ic_1 T + 3c_2 T^2 + im T^3 +
ih_0 S - 3a ST - \sum_{k=0}^3 (ih_k + b_kT) U_k\ ,
\label{sup2}
\eeq 
If the fluxes $h_k$ and $b_k$ are independent of $k$,
we can also set $U_1=U_2=U_3=U$.

Given the fluxes leading to (\ref{sup2}), the tadpole condition
(\ref{tadxxx}) becomes 
\beq 
\sum_a N_a n_a^1 n_a^2 n_a^3 + \frac12(h_0
m - 3a c_2) = 16 \ .
\label{tadxxxiso}
\eeq 
On the other hand, 
\beqa 
\sum_a N_a n_a^1 m_a^2 m_a^3 + \frac12 (h_1 m + b_1 c_2) & = & 0  \  ,
\nonumber \\[0.2cm] 
\sum_a N_a m_a^1 n_a^2 m_a^3 + \frac12(h_2 m + b_2 c_2) & = & 0  \  ,
\label{tadodhiso} \\[0.2cm]
\sum_a N_a m_a^1 m_a^2 n_a^3 + \frac12(h_3 m + b_3 c_2) & = & 0 \ . \nonumber 
\eeqa 
In the next section we will see that to obtain a
minimum of the moduli scalar potential the fluxes must satisfy some
relations that will in particular fix the sign of the flux
contributions to the tadpoles.


\section{Vacuum structure in IIA orientifolds with fluxes}
\label{sec:vac}

In this section we analyze the moduli scalar potential induced by the
fluxes. For the K\"ahler potential we assume the usual tree-level
result displayed in (\ref{kpot}).  The scalar potential is then simply 
\beq
V= e^K \big\{ \sum_{\Phi=S,T_i, U_i} \, (\Phi + \Phi^*)^2 |D_\Phi W|^2
- 3|W|^2 \big\} \ ,
\label{sv}
\eeq 
where $D_\Phi W = \partial_\Phi W + W \partial_\Phi K$ and $W$ is given in
eq.(\ref{wa}).
 We want
to look for solutions of $\partial V/\partial \Phi = 0$.  A well known
and easy to show general result is that the simpler unbroken susy
conditions $D_\Phi W=0$ imply a minimum.

As it happens in the IIB case \cite{kst, lrs1}, it is quite complicated to 
perform a complete analysis of all possible minima induced by the flux
superpotential. We have however explored several possibilities 
including $\neq1$ supersymmetric Minkowski vacua, no-scale models in
Minkowski, $\neq1$ supersymmetric AdS vacua and non-supersymmetric AdS vacua.
Before providing specific details let us make some general comments 
about these vacua and their comparison to IIB mirror orientifolds
discussed in \cite{kst, lrs1}. In type IIB the flux-induced
superpotential only depends on the axion-dilaton and complex
structure fields, and all K\"ahler moduli remain undetermined.
In the IIA case at hand all moduli appear in the 
superpotential and in principle may be fixed. There are however 
classes of type IIB minima whose mirror IIA dual is not included
among our vacua. In particular, those arising when the IIB  
superpotential is linear in the axion-dilaton $\tau$ and at the same time 
quadratic/cubic in the complex structure fields. In fact, most of the examples 
in refs.~\cite{kst, lrs1} belong to this class.
A naive IIB/IIA mirror symmetry would suggest terms in the superpotential linear
in $S$ and quadratic/cubic in the K\"ahler moduli which are not present in
our superpotential (\ref{wa}).  Otherwise, the IIA options for
fluxes are richer and lead to many new possibilities, not only in the number 
of determined moduli but also in the contribution of fluxes to the
different RR tadpoles, as we discuss below.

Constraining to the dependence on the 7 moduli,
$S$ and the diagonal $U_i,T_i$, in general one finds that to get (non-runaway) 
minima the superpotential has to depend at least on four moduli.
Note also that the fields $S$ and $U_i$ appear in a similar
(linear) way in the superpotential so that, e.g. given a 
superpotential depending on $S$, one can obtain another model 
replacing $S\rightarrow U_i$ and properly relabelling 
the fluxes. Compared to the original model, the new one so derived has 
in general different contributions to the RR tadpoles.
The same is true for the $T_i$ moduli in the $m=0$ case. 
One can obtain new models replacing
$T_i\rightarrow U_i$ and exchanging appropriately the fluxes.
This will be illustrated below. 

In this section we then use the flux-induced superpotential
to study \neq1 and \neq0 type IIA vacua. In appendix A we will
also discuss how \neq1 vacua in the presence 
of metric fluxes can be analyzed in terms of compactifications
on the twisted torus seen as a manifold with $SU(3)$ structure.

Let us now discuss the different types of vacua in turn. 
To describe the
results we will denote $\re T_i \equiv t_i$, $\im T_i \equiv v_i$, $\re S
\equiv s$ and $\re U_i \equiv u_i$.

\subsection{ Supersymmetric Minkowski vacua}
\label{ssminkowski}

Supersymmetric Minkowski vacua have $\langle D_\Phi W\rangle =0$ with
$\langle V\rangle=0$ so that the cosmological constant vanishes.  
Clearly, (\ref{sv}) then implies $\langle W\rangle =0$ and
the supersymmetry condition reduces to $\langle\partial W/\partial
\Phi\rangle=0$. To simplify notation, in the following we stop
writing vevs explicitly, it should be obvious when quantities are
meant to be evaluated at the extremum.

When $W$ depends on only three fields it suffices to analyze
$W=W_K(T_1,T_2,T_3)$, purely due to RR fluxes. Other cases can be
easily translated.  The general solution of $D_{T_i}W_K=0$ was
found in \cite{llm} in relation to BPS black holes, and applied in
\cite{cklt} in the context of type IIA vacua with fluxes. If
$W_K=0$ there are only pathological solutions in which e.g. $t_1=0$.

Going to superpotentials depending on four moduli or more it is 
easy to find models with $\neq1$ supersymmetry in Minkowski space. 
One can obtain  examples of this kind of minima with the
superpotential depending on  a) one $S/U_i$ modulus and three $T_i$; b) two 
$S/U_i$ fields and two $T_i$ and c) two  $S/U_i$  and three $T_i$.
We describe these cases below.
We have not found any models with superpotentials depending 
on more than 5 fields. In all the examples we find we need the 
presence of metric fluxes and in addition $m=0$.
 In all cases there is only a partial fixing of moduli, 
 our examples fixing  at most 3 complex  linear combinations
of moduli. 
The fluxes in these examples contribute to the RR tadpoles 
with the same sign as D6-branes do. However they may contribute 
in any of the four RR tadpole directions, depending on each case.
This is explained below.

\subsubsection{Superpotentials depending on four moduli}

With four fields it is enough to study in detail $W=W(S,T_1,T_2,T_3)$,
c.f. (\ref{wast}), independent of the $U_i$.  Other cases,
e.g. $W(S,U_1,T_2,T_3)$, can be mapped into this by renaming fields and
choosing parameters appropriately.  We look for solutions of $W=0$,
$\partial_SW=0$ and $\partial_{T_i}W=0$.  If there are no metric
fluxes we find that $W$ must vanish identically to avoid $t_i=0$.  If
instead $a_i \not=0$ there are solutions provided $m=0$.
Moreover, taking the real part of $\partial_\Phi W=0$ gives four homogeneous
equations $a_1 t_1+a_2 t_2 + a_3 t_3=0$ and 
$a_i s+ q_j t_k + q_k t_j=0$, $i\not=j\not=k$. To have $s, \, t_i \not=0$
the determinant of this system must vanish, this is
\beq (a_1
q_1 + a_2q_2 - a_3 q_3)^2 = 4 a_1q_1 a_2 q_2 \ .
\label{mini1c1}
\eeq 
Let us discuss now some particular cases.

\noindent
{\it i) Example SM-1}

\noindent
We can for example take $a_1=0$ which requires $q_1=0$ to avoid
$t_2,\ t_3 =0$. The superpotential has the general form
\beq
W\ =\ e_0+ih_0S+i\sum_ie_iT_i-S(a_2T_2+a_3T_3)-T_1(q_2T_3+q_3T_2)  \  .
\label{MS1}
\eeq 
It must be $a_2q_2 =a_3 q_3$ and for consistency  one  also
needs  $e_2a_3=e_3 a_2$, $h_0 q_2=a_3 e_1$, and $h_0 e_2=e_0 a_2$.
Note that these conditions are easily satisfied, e.g. 
for $h_0=e_I=0$. Neither the imaginary nor the real
parts of the moduli are fully determined, only 
\beq 
h_0=a_2 v_2 + a_3
v_3 \quad ; \quad e_2=a_2 \im S + q_3 v_1 \quad ; \quad t_3=-\frac{q_3
t_2}{q_2} \quad ; \quad s=-\frac{q_3 t_1}{a_2} \ .
\label{mini1c2}
\eeq 
For $s, t_i > 0$ we must have $q_2 q_3 < 0$ and $a_2 q_2>0$.
Now, the flux term in the (\ref{tadxxx}) RR tadpole
is equal to $2a_2q_2$ and is therefore positive.

\noindent
{\it ii) Example SM-2}

\noindent
The above  example with $a_1=0$ can be used to analyze vacua in which
we replace one K\"ahler, say $T_1$, by a complex structure modulus, 
say $U_1$. The $W(S,U_1,T_2,T_3)$ superpotential is now
\beq
W = -T_2(a_2S+b_{21}U_1) - T_3(a_3S+b_{31}U_1)  + e_0 + ih_0S\ -
ih_1U_1  + i e_2T_2  + ie_3T_3  \  .
\label{SM2}
\eeq
This is clearly equivalent to (\ref{MS1}) after renaming $T_1 \to
U_1$, $e_1 \to -h_1$, $q_2 \to b_{31}$ and $q_3 \to b_{21}$.
The physics is however different. In particular, since all $q_i=0$
and $m=0$, these fluxes do not contribute at all to the RR tadpoles.
Thus, this is an example in which one can fix moduli without
affecting tadpoles, the fluxes are essentially unconstrained, except 
from the consistency conditions mentioned above.
We will see later that there are $\neq1$ supersymmetric AdS vacua in 
which one can fix all moduli without any restriction from RR tadpoles.

\noindent
{\it iii) Example SM-3}

\noindent
Another simple solution of
(\ref{mini1c1}) is to take $a_2=a_3=-a_1/2$, and $q_2=q_3=-q_1/2$. Further
selecting $h_0=0$ and $e_I=0$ one has a superpotential 
\beq
W\ =\ -a_1 ST_1 - q_1 T_2 T_3 + \oh (T_2 + T_3)(a_1 S + q_1 T_1)  \  .
\label{SM3}
\eeq
One obtains a supersymmetric Minkowski minimum
with $T_1=T_2=T_3=a_1 S/q_1$. The fluxes contribute 
$(3a_1q_1/2) > 0$ to the (\ref{tadxxx}) RR tadpoles.

\subsubsection{Superpotentials depending on five or more moduli}

With five fields there are solutions of $W=0$ and $\partial W/\partial
\Phi=0$. This is our next example.

\noindent
{\it iv) Example SM-4}

\noindent
Consider the superpotential
\beq W= 
-(q_3T_2 + q_2 T_3)T_1-(b_{22}T_2 + b_{32} T_3)U_2- (b_{23}T_2 + b_{33} T_3)U_3 \ .
\label{w5f}
\eeq 
Observe that the non-zero $b_{ij}$ trivially satisfy the Jacobi
identities (\ref{jacb}). If the fluxes satisfy $q_2 b_{22} = q_3
b_{32}$, $q_2 b_{23} = q_3 b_{33}$ and $q_2 q_3 < 0$, there is a
solution with $|q_2| t_3 = |q_3| t_2$. There is also a relation $-q_3
t_1= b_{22} u_2 + b_{23} u_3$.  To have $t_1 > 0$ for any $u_2$, $u_3
> 0$, we need $q_2 b_{22} > 0$ and $q_2 b_{23} >0$. Hence, the flux
piece in (\ref{tadodh}) is negative (same as D6-branes).
In this particular case fluxes contribute $2b_{22}q_2$ to the 
last two RR tadpoles in (\ref{tadodh}).

As we have said,
for the given class of $W$'s, in which the metric fluxes must satisfy
the Jacobi identities (\ref{jacb}), there cannot be supersymmetric
Minkowski solutions when $W$ depends on more than five fields.  To see
this, first observe that without loss of generality we can always take
the three $T_i$ to be among the fields in $W$. Next, from
(\ref{tildew}) we see that $\partial W/\partial \wt{U}_K=0$ implies
$A_{IJ} \wt{T}_I=0$, and taking real part, $A_{iJ} t_i\equiv \ca_{Ji}
t_i=0$, where $\mathsmaller{J}$ takes three or four values.  Note that
the $t_i$ correspond to the kernel of the metric flux matrix $\ca$.
Thus to have solutions with $t_i \not=0$, ${\rm rank}\, \ca \le 2$.
After using the Jacobi identities one is left with ${\rm rank}\,
\ca=1$. One can then check that the number of fields in $W$ can be
at most five.  However, we will see that for $W$ depending on all seven moduli, 
there are supersymmetric AdS minima in which $D_\Phi W=0$ but $W\not=0$.

As all examples so far show, fluxes allowing 
$\neq1 $ supersymmetric Minkowski vacua only fix the moduli partially.
Recall that in the type IIB case there are supersymmetric Minkowski
minima with all complex structure fields fixed (but not the
K\"ahler moduli). This is because, as already mentioned, 
in IIB there are extra superpotential couplings $\tau \tau_i \tau_j$ or 
$\tau \tau_1 \tau_2 \tau_3$,
whose naive IIA mirror $S T_i T_j$, $ST_1 T_2 T_3$ we do not have.
This situation changes when $W$ depends on 
the seven moduli. As we will see, supersymmetric AdS
minima with all real moduli determined can then occur.
This will be discussed in sections \ref{ssadsnf} and \ref{ssadsf}.

\subsection {No-scale models in Minkowski space}
\label{ssnoscale}

As `no-scale' we distinguish models  
in which the superpotential is independent of three of the moduli, so that the 
form (\ref{kpot}) of the K\"ahler potential guarantees a cancellation of the
cosmological constant \cite{Cremmer}. In fact, the scalar potential is 
positive definite and it is minimized with respect to all fields when 
$D_{\Phi^\prime} W(\Phi^\prime)=0$. Since in general $W\not=0$, supersymmetry 
is broken by the F-terms of the fields not appearing in $W$. 

One easily finds no-scale minima with superpotentials depending 
on the moduli sets $\{U_I,T_1,T_2,T_3\}$ or $\{U_I,U_j,T_k,T_\ell\}$,
with $U_I=S,U_i$. There are no minima coming from 
superpotentials of the form $W(U_I,U_j,U_k,T_\ell)$ or 
$W(S,U_1,U_2,U_3)$.
Unlike the previous case with $\neq1$ supersymmetry, one can find 
no-scale models with and without metric fluxes.
In the latter case one necessarily has $m\not= 0$. 
As in the $\neq1$ supersymmetric examples, the moduli are
typically only partially fixed and the fluxes contribute to
RR tadpoles like D6-branes.
In particular, if $W$ only depends on $S,T_i$, the fluxes
only contribute to the (\ref{tadxxx}) RR tadpoles.
If it depends on one or two  $U_i$'s, fluxes 
may contribute to other RR tadpoles, but always positively.
We can consider a  superpotential $W=W(S,T_1,T_2,T_3)$ as
generic, since one can always replace $S$ or one or two
$T_i$'s by $U_i$'s if appropriate fluxes are also exchanged.

Our task is then to look for solutions of
$D_SW=0$ and $D_{T_i}W=0$, without imposing $W=0$,
starting with the $W(S,T_1, T_2, T_3)$ given in eq.(\ref{wast}).
We remind the reader that this superpotential 
may be obtained performing T-dualities
on the superpotential $\cw$ of type IIB generated by certain $\ov{\cf}_3$
and $\ov{\ch}_3$ fluxes. Now, in the type IIB case, solving $\D_\tau \cw=0$,
$\D_{\tau_i} \cw=0$, is equivalent to demanding that the flux
$\ov{\cg_3}=(\ov{\cf}_3 -\tau \ov{\ch}_3)$ be imaginary self dual (ISD) \cite{gkp}. 
Indeed, e.g. the conditions (\ref{min3c1}), (\ref{min3c2}) and (\ref{min3c3}) below 
amount to the ISD requirement.

\subsubsection{Examples with no metric fluxes}

\noindent
{\it i) Example NS-1}

\noindent
In the case with no metric fluxes 
one can check that there are minima 
only if $m \not=0$, $h_0 \not=0$, and $\g_1=\g_2=\g_3=0$, where the
$\g_i$ are the combination of RR fluxes 
\beq 
\g_i= m e_i + q_j q_k
\quad ; \quad i \not= j \not = k \ .
\label{allg}
\eeq
The superpotential has the form  
\beq
W = e_0+ih_0S+i\sum_ie_iT_i-q_1T_2T_3-q_2T_1T_3-q_3T_1T_2+imT_1T_2T_3  \  . 
\label{NS1}
\eeq
For the moduli we determine $v_i = -q_i/m$, $\im S=(e_0 m^2 -
q_1q_2q_3)/h_0m^2$, whereas for the real parts $h_0 s = m t_1 t_2
t_3$. Hence, $h_0 m > 0$ and again the flux contribution to tadpoles
is positive. We also find that at the minimum the superpotential is
$W_0=2 i h_0s$ and the gravitino mass is $m^2_{3/2}= h_0 m/32 u_1 u_2
u_3$.
Note that this class of background is mirror to an
analogous class of no-scale models in type IIB discussed in
\cite{ms}.

\subsubsection{Examples with metric fluxes}

With a superpotential of type $W(S,T_1,T_2,T_3)$ this amounts to 
having $a_i\not=0$ for some $i$. Let us consider some simple models 
giving rise to a no-scale structure.

\noindent
{\it ii) Example NS-2}

\noindent
Let us  study first the case in which $m=0$ and 
one of the $a_i$ vanishes, say $a_1=0$. We will then be able to
translate the results for, say $W(S,U_1,T_2,T_3)$ if we wish.  
The superpotential has the form 
\beq
W\ =\ e_0+ih_0S+i\sum_ie_iT_i-S(a_2T_2+a_3T_3)-q_1T_2T_3-q_2T_1T_3-q_3T_1T_2  \  .
\label{NS2}
\eeq
If $m=0$, then $t_2,
t_3 \not=0$ require $q_1=0$. One finds that the 
 axions are then fully determined to be 
\beq 
\im S=\frac{e_2 q_2-e_3q_3}{q_2 a_2 - q_3 a_3} \ ; \
v_1=\frac{e_3 a_2 - e_2 a_3}{q_2 a_2 - q_3 a_3} \ ; \ v_2=\frac{h_0
q_2 - e_1 a_3}{q_2 a_2 - q_3 a_3} \ ; \ v_3=\frac{e_1 a_3 - h_0
q_3}{q_2 a_2 - q_3 a_3} \ .
\label{min2c1}
\eeq 
There is a further relation $e_0=h_0 \im S + e_1 v_1$. The real
parts verify $q_2 t_1 t_3 = a_2 s t_2$ and $q_3 t_1 t_2 = a_3 s
t_3$. Thus we must have $a_2 q_2 > 0$ and $a_3 q_3 >0$,
indicating a positive  contribution to the (\ref{tadxxx}) RR tadpoles. 
On the other hand  only
pairwise ratios of real moduli are fixed, namely 
\beq 
s^2=\frac{q_2q_3}{a_2 a_3}t_1^2
\quad ; \quad t_3^2=\frac{a_2 q_3}{a_3q_2} t_2^2 \ .
\label{min2c2}
\eeq 
At the minimum, $W_0=-2s(a_2 t_2 + a_3 t_3)$. 

In a variant of this model one  can further set
$a_2=0$ and for consistency $q_2=0$. The imaginary parts are obtained
substituting these values in (\ref{min2c1}). For the real parts it
only follows that $a_3 s t_3 = q_3 t_1 t_2$.  
In yet another variant one can set  $a_2 q_2 =a_3 q_3$.
The imaginary parts are then given as in (\ref{mini1c2}), while
$a_3 t_3 = \pm a_2 t_2$ and $a_2 s = \pm q_3 t_1$. This allows either
$a_2 a_3 >0$ or $a_2 a_3 <0$ (so we could further impose $W_0=0$ as in the model SM-1).

\noindent
{\it iii) Example NS-3}

\noindent
Let us now consider the case with $m\not=0$, still with $a_1=0$.
One finds that in order to have a solution the fluxes must
verify 
\beq 
\g_2=\frac{a_2 \g_3}{a_3} \quad ; \quad h_0 \g_3=a_3(e_1
q_1 + m e_0) \ .
\label{min3c1}
\eeq 
For the imaginary parts we obtain 
\beq 
\im S=\frac{me_0 + q_1
e_1}{m h_0} \ ; \ v_1=-\frac{q_1}{m} \ ; \ v_2=-\frac{q_2}{m} +
\frac{a_2 s t_2}{m t_1 t_3} \ ; \ v_3=-\frac{q_3}{m} + \frac{a_3 s
t_3}{m t_1 t_2} \ .
\label{min3c2}
\eeq 
The real parts instead satisfy 
\beq 
a_2 a_3 s^2 = \g_1 t_1^2
\quad ; \quad m^2 t_1^2 t_2^2 t_3^2 + a_2^2 s^2 t_2^2 + a_3^2 s^2
t_3^2 = (h_0 m + a_2 q_2 + a_3 q_3) s t_1 t_2 t_3 \ .
\label{min3c3}
\eeq 
This shows that $(h_0 m + a_2 q_2 + a_3 q_3) > 0$ and hence the
flux contribution to tadpoles is again positive. Notice that the above
solution simplifies upon taking $a_2=0$ which is consistent if
$\g_1=\g_2=0$. In this case 
\beq 
t_3 = \frac{(h_0 m + q_3 a_3) s t_1
t_2}{(a_3 s)^2 + (m t_1 t_2)^2} \ .
\label{t3v}
\eeq 
We also find that at the mimimum 
\beq 
W_0 = -\frac{ 2(h_0 m + q_3 a_3) s t_1 t_2} {a_3 s + i m t_1 t_2} \ .
\label{wval}
\eeq 
The gravitino mass turns out to be 
\beq 
m^2_{3/2}= \frac{ (h_0 m + q_3 a_3)} {32 u_1 u_2 u_3} \ .
\label{mgrav}
\eeq 
As expected for a no-scale model it only depends on the $u_i$.

\noindent
{\it iv) Example NS-4}

\noindent
Consider now an isotropic setup with
 $a_1=a_2=a_3=a$ and $q_1=q_2=q_3=q$. It is
possible to obtain vacua with the four moduli fixed \cite{dkpz}. At
the minimum, $\re T_k = t$, $\im T_k = v$.  After taking $e_I=0$ we
find the equations 
\beqa 
v[2 a m v^2 - (h_0m -3 aq)v - 2 h_0q] & = & 0
\ , \nonumber \\[0.2cm] (h_0-3av)(q+mv) & = & a m t^2 \ .
\label{vtiso}
\eeqa 
The dilaton is given by $aS=(q+mv)T^* - iqv$. There is a
solution with $v=0$ which, since then $T=\sqrt{h_0 q/m a}$, $aS=qT$,
can occur only if $h_0m >0$ and $aq > 0$. A solution with $v\not=0$
can happen if $(aq/h_0m) < -1$ and $h_0m <0$, or if $-1/9 < (aq/h_0m)
< 0$ and $h_0m >0$. In all cases the flux contribution to the RR
tadpole is positive.

\subsection{AdS vacua without metric fluxes}
\label{ssadsnf}

We now consider the superpotential, depending on all seven moduli,
given by the sum of the two terms in (\ref{sup0}). As we will see,
as long as $m\not=0$, $T_i$, $s$ and $u_i$  can be fixed in AdS, but only a linear 
combination of the axions $\im S$ and $\im U_i$ is determined. This
can occur both for broken and unbroken supersymmetry but in all cases 
we prove that the contribution of fluxes to RR tadpoles is always positive.

We find that in the absence of  metric fluxes
the vacuum structure is determined by the
combination of fluxes $\g_i$, c.f. (\ref{allg}). In particular, $\g_i < 0$
is required to solve $D_{T_i}W=0$, $D_S W=0$, and $D_{U_i}
W=0$ in the supersymmetric case.   Then, the K\"ahler axions are
fixed as $v_i=-q_i/m$, whereas for the other axions
\beq
h_0\im S - \sum_i h_i\im U_i = \frac1{m^2}[e_0 m^2 - q_1q_2q_3 + \sum_i q_i \g_i]
\ .
\label{imflat}
\eeq
The real parts are instead determined to be
\beq
t_1 = \sqrt{\frac{5 |\g_2 \g_3|}{3 m^2 |\g_1|}} \quad ; \quad t_2 =
\frac{\g_1 t_1}{\g_2} \quad ; \quad t_3 = \frac{\g_1 t_1}{\g_3} \quad
; \quad s = -\frac{2\g_1 t_1}{3 mh_0} \quad ; \quad u_i = \frac{2\g_1
t_1}{3mh_i} \ .
\label{isosusy}
\eeq
Observe that in order to have $s > 0$, $u_k >0$, it must be that
$mh_0 > 0$ whereas $mh_k < 0$.  Hence, the flux contribution to the
tadpoles is positive in (\ref{tadxxx}) and negative in (\ref{tadodh}).
Note that in the present case only $m$, $h_0$ and $h_k$ are restricted by RR tadpole
conditions while the fluxes $e_0$, $c_1$ and $c_2$ are essentially unconstrained.
This will allow us to find minima at arbitrarily large volume and small dilaton,
see below.
Type IIA supersymmetric AdS vacua without metric fluxes have been
recently addressed in \cite{DeWolfe}. We obtain similar results.

To go beyond supersymmetric minima and find all solutions of $\partial
V/\partial \Phi=0$ we will analyze the case $T_1=T_2=T_3=T$, so that  
$W$ is given in (\ref{sup2}) with $b_k=a=0$.  The vacuum structure now
depends on $\gamma=m c_1 + c_2^2$.  In particular, there exists a supersymmetric
AdS minimum only if $\g < 0$.  We also find that necessarily $m h_0 >
0$ and $m h_k < 0$.  Therefore, the flux contribution to the tadpoles
is positive in (\ref{tadxxxiso}) and negative in (\ref{tadodhiso}).  

To summarize the results we use the shorthand $\re T=t$ and $\im T=v$.
The extremum only fixes one linear combination of the imaginary parts
of dilaton and complex structure fields which is given by
\beq
h_0\im S -\sum_{k=1}^3 h_k\im U_k = e_0 - 3c_1v -3c_2 v^2 + m v^3 \ .
\label{zval}
\eeq
There are two branches for $v$, namely,
\beq
v_s=\frac{c_2}{m} \quad ; \quad
v_{ns}=\frac{c_2 \pm \sqrt{\g -m^2 t^2/2}}{m} \ .
\label{ysns}
\eeq
For each value of $v$ there are various sub-branches according to
the relation among the real parts of $S$ and the $U_k$. From now on we
just look at solutions with $h_1 u_1=h_2 u_2= h_3 u_3$. In this case
there are two sub-branches characterized by
\beqa
({\rm I}) & : & h_k u_k = -h_0 s \quad ; \quad k=1,2,3 \nonumber \\[0.2cm]
({\rm II}) & : & h_k u_k = h_0 s - m t^3 \ .
\label{subus}
\eeqa
In the $v_s$ sub-branch I,
\beq
m^2 t^2 = \pm \frac53\g \quad ;
\quad h_0 s=\ds{\frac2{5} m t^3} \quad ; \quad
\Lambda_s=-\ds{\frac{\g^2}{24 m^2 s u_1 u_2 u_3 t}} \ .
\label{branchss}
\eeq
For $\g < 0$ this is the AdS supersymmetric minimum. For $\g > 0$ it is a
non-supersymmetric AdS extremum with same data for the moduli and the
cosmological constant. In the $v_s$ sub-branch II,
\beq
m^2 t^2 = \pm \frac{5}{\sqrt6}\g \quad ; \quad h_0 s=\ds{\frac4{5} m t^3} \quad ;
\quad \ds{\frac{\Lambda}{\,
\Lambda_s}=\frac{32}{27}\left(\frac69\right)^{1/4}} \sim 1.071 \ .   
\label{branchsns}
\eeq
Both $\g < 0$ and $\g > 0$ are allowed. In either case it is a
non-supersymmetric AdS extremum.

The $v_{ns}$ branch can occur only if $\g > 0$. There are two
non-supersymmetric AdS sub-branches according to (\ref{subus}).
Their data are
\beqa
({\rm I}) \! & : & \! m^2 t^2 =
\frac43\g \quad ; \quad h_0 s=\ds{\frac{2\g t}{3m}} \quad ; \quad
\ds{\frac{\Lambda}{\, \Lambda_s}=\frac{25\sqrt5}{48}} \sim 1.165 \ ,
\nonumber \\[0.2cm]
{} & & \label{branchnsII} \\[0.2cm] ({\rm II})\!
& : & \!  m^2 t^2 = \frac{196}{99}\g \quad ; \quad h_0
s=\ds{\frac{14\g t}{9m}} \quad ; \quad \ds{\frac{\Lambda}{\,
\Lambda_s}=\frac{11^4 5^2 3^2\sqrt{55}}{2^4 7^7 \sqrt3}} \sim 1.070  
\ . \nonumber
\eeqa
This ends our list of non-supersymmetric AdS vacua.

Note that in all these examples without metric fluxes
the fixed moduli scale with respect to the RR 4-form
and 2-form  fluxes $c_1$ and $c_2$ as
\beq
t\ \simeq \  s^{1/3} \ \simeq \ u_k^{1/3} \ \simeq  \ \gamma^{1/2}  \
\simeq
\ c_1^{1/2} \ ,\ c_2   \  ,
\label{scalings2}
\eeq
for large fluxes. Thus the compactification volume may be arbitrarily large
for large $c_1$ and/or $c_2$. For large fluxes, the four- and ten-dimensional dilatons 
behave as
\beq
e^{\phi_4} \ \simeq \ c_1^{-3/2} \ ,\ c_2^{-3} \ \ ;\ \
e^{\phi} \ \simeq \ c_1^{-3/4} \ ,\ c_2^{-3/2}
\label{dili}
\eeq
so that the vacua lie in a perturbative regime for sufficiently
large RR 4-form and/or 2-form fluxes.
Concerning the cosmological constant, one can check that for large 
fluxes $c_1$ and $c_2$ it scales as 
\beq
V_0\ \simeq \ -\ \gamma^{-9/2} \ \simeq \ -c_1^{-9/2}\ , \ -c_2^{-9} \ .
\label{cosmolon}
\eeq
Thus, for large $c_1$/$c_2$ the c.c. goes with the string dilaton like
$e^{6\phi}$.
The density of RR fluxes is also suppressed. As pointed out in
\cite{DeWolfe}, to compute this density a factor of $g_s=e^\phi$
must be included. Then, the flux density of $\ov{F}_4$
($\ov{F}_2$) behaves like $c_1^{-3/2}$ ($c_2^{-3}$) for
large $c_1$ ($c_2$) fluxes.

\subsection{Supersymmetric AdS vacua with metric fluxes}
\label{ssadsf}

We finally treat the full superpotential given in (\ref{wa}). 
We will see that all real moduli may be fixed at the 
minimum. Concerning the imaginary parts, some linear combinations of
$\im S$, $\im U_i$ remain massless but, as we will discuss in the 
next section, in the presence of D6-branes those massless fields
are in fact necessary for certain 
(potentially anomalous) brane $U(1)$ fields to get
a St\"uckelberg mass. We will also see that to get these minima 
certain discrete relationships among the fluxes must be fulfilled.

There are two classes of models depending 
on whether $m=0$ or not. In the former case one finds 
that fluxes in general contribute to all RR tadpole directions with
a sign which is {\it opposite } to that of D6-branes. This is 
important since it offers an alternative to orientifold planes 
to cancel RR tadpoles. In the second case with
$m\not= 0$ one finds the interesting result that, depending
on different flux choices, always including metric fluxes,
the sign of the contribution to RR tadpoles may be arbitrary and
the net contribution may vanish. In the latter case one has 
a cancellation of a positive RR-NS contribution $h_0m$ with 
a RR-metric flux contribution of type $a_iq_i$. This is interesting
because in this class of backgrounds all real moduli
are determined but the fluxes are unconstrained by 
RR tadpole cancellation conditions.

We will examine the case $T_k=T$ and look for supersymmetric
minima for any $m$. From $D_{U_k}W=0$ and $D_SW=0$, with $W$ given in
(\ref{sup2}), we find
\beq
3a s = b_k u_k \ .
\label{fixsu}
\eeq
Hence, $a$ and $b_k$ must be both non-zero and of the same sign.
Moreover, there are consistency conditions
\beq
3h_k a + h_0 b_k = 0
\quad ; \quad k=1,2,3 \ .
\label{finetune}
\eeq
Therefore, either both $h_0$ and $h_k$ vanish or both are non-zero and of
opposite sign. These conditions do not involve the moduli so at most  
we will have five equations for six unknowns, i.e. we will have at
least one flat direction for the supersymmetric minima. In fact, only
a combination of complex structure axions is fixed as
\beq
3a \im S +\sum_{k=1}^3 b_k \im U_k = 3c_1 + \frac{3c_2}{a}(3h_0
- 7a v) - \frac{3m}{a} v(3h_0 - 8a v) \ .
\label{flatsu}
\eeq
If $h_0, h_k \not=0$ using (\ref{finetune}) we can write the fixed axion
combination
as $h_0 \im S -\sum_k h_k \im U_k$.
We also find  
\beq a s = 2 t (c_2 - m v) \ .
\label{fixres}
\eeq
Except for some axion directions,  we have thus determined all the
moduli in terms of $T$ which is found from the remaining
equations. The solution depends on whether $m$ is different from zero
or not. We now specify to these two possibilities.

\noindent
{\it i) $m=0$. Examples  AdS-1}

\noindent
When $m=0$ we find the simple results
\beq
v=\frac{h_0}{3a} \quad ; \quad 9c_2 t^2 = e_0 - \frac{h_0 c_1}{a} -
\frac{h_0^2 c_2}{3a^2} \ .
\label{valtm}
\eeq
At the minimum, $W_0=-12 c_2 t^2$ and the cosmological constant turns out to be
\beq
V_0= \Lambda = - \frac{a b_1 b_2 b_3}{128 \, c_2^2 \, t^3} \  ,
\label{cosmo1}
\eeq
where $t$ is given in (\ref{valtm}). It is important to
notice that (\ref{fixres}) fixes $c_2 a > 0$ so that in the
supersymmetric minima with $m=0$ the metric fluxes give a negative
contribution to the tadpole in (\ref{tadxxxiso}). Similarly, $c_2 b_k > 0$
and the flux contribution to the tadpoles in (\ref{tadodhiso}) is positive.
This is the result we advertised. 

Let us now check whether 
we have enough freedom to locate all moduli at large volume and small 
dilaton so that one can trust the effective 4-dimensional field theory
approximation being used. The fluxes of type  $a,b_k$ and $c_2$
are constrained by the RR tadpole cancellation conditions and by 
the extra conditions (\ref{finetune}). The values of $h_0$ and $h_k$ 
are constrained by the latter but in principle both may be large 
as long as $h_0/h_k=-3a/b_k$. On the other hand, 
the fluxes  of 
the RR 6-form ($e_0$) and 4-form ($c_1$) are unconstrained and may be
arbitrarily large. Note then that for large $e_0$ and $c_1$ the 
moduli fields behave all like
\beq
t\ \simeq \  s \ \simeq \ u_k \ \simeq e_0^{1/2} \ ,\ c_1^{1/2}  \  .
\label{scalings1}
\eeq
In order for  our vacuum to remain in a perturbative 
regime we would like to have small values for the 4-dimensional 
coupling $e^{\phi_4}$ and the 10-dimensional string coupling 
$e^{\phi}$. They are found to be 
\beqa
e^{\phi_4} \ &=& \ (su_1u_2u_3)^{-1/4} \ =\ 
t^{-1}  \frac {(a b_1 b_2 b_3)^{1/4}}{2 \cdot 3^{3/4} c_2}  \ ,
\nonumber \\[0.2cm]
 e^{\phi} \ &=& \  e^{\phi_4}t^{3/2}\ =\  t^{1/2} 
\frac {(a b_1 b_2 b_3)^{1/4}}{2 \cdot 3^{3/4}c_2}  \ .
\label{maldilaton}
\eeqa
We thus see that for large $t$ (which may be obtained e.g. with
a large 6-form flux $e_0$) the 4-dimensional dilaton is small.
However the string dilaton grows with $t$. Only by appropriately
choosing the fluxes, i.e. with large $c_2$ one can perhaps
maintain it under control. On the other hand such fluxes are
in general very much constrained by the RR tadpole conditions
so it seems difficult having small string dilaton and large
volume at the same time. We will see however that in the case with
$m\not=0$ one can easily stabilize the moduli in the perturbative regime.

\noindent
{\it ii) $m\not=0$. Examples  AdS-2}

\noindent
To deal with $m \not=0$ it is convenient to introduce a new variable
for $\im T$. If $h_0 \not=0$ we use
\beq
v= (\lambda + \lambda_0)\frac{h_0}{3a} \quad ; \quad 
\lambda_0=\frac{3c_2 a}{m h_0} \ .
\label{newv}
\eeq
The value of $\lambda$ follows from the cubic equation
\beq 
160\lambda^3 + 186(\lambda_0-1)\lambda^2 + 27(\lambda_0-1)^2\lambda+
\lambda_0^2(\lambda_0-3) + \frac{27a^2}{m h_0^3}(e_0 a - c_1 h_0) = 0 \ .
\label{cubice}
\eeq
Clearly, we need a real solution for $\lambda$ and moreover, such that
$\lambda(\lambda + \lambda_0-1) > 0$ because
now $t$ is determined from
\beq
3a^2t^2= 5h_0^2 \lambda(\lambda + \lambda_0-1) \ .
\label{newt}
\eeq
Notice also that (\ref{fixres}) takes the form $3a^2 s=-2 h_0 m \lambda t$.
For the cosmological constant we find
\beq
V_0= - \frac{a b_1 b_2 b_3 \lambda_0^2 (16 \lambda + \lambda_0 -1)}
{1920 \, c_2^2 \, t^3 \lambda^3} \  ,
\label{cosmo2}
\eeq
where $\lambda$ is the appropriate solution of (\ref{cubice}) and $t$ 
is given in (\ref{newt}).

There is a variety of cases depending on the values of the different fluxes.
One of the interesting features when $m\not=0$ is that the contribution
to the RR tadpoles may have either sign and even vanish. In fact,
the flux-induced tadpoles in (\ref{tadxxxiso}) and (\ref{tadodhiso})
are respectively $\frac12 h_0m(1-\lambda_0)$
and $\frac12 h_km(1-\lambda_0)$. Thus, the flux tadpoles vanish at the
special  value  $\lambda_0=1$. This is important, as we mentioned above,
since it allows to fix the moduli without any constraint from
RR tadpole cancellations.

To analyze the equations that determine $\lambda$  and $t$ we can proceed in various ways. 
We could choose for example $e_0$ and $c_1$ and study the allowed values
of $\lambda_0$. For instance, with $e_0=c_1=0$, we find that to have solutions for
$\lambda$ with acceptable $t$ necessarily $\frac13 < \lambda_0 < 3$. We
also need $m h_0 < 0$ and $m h_k > 0$ so that $s > 0$ and $u_k > 0$.
The special value $\lambda_0=1$ at which tadpoles vanish is allowed and 
leads in turn to 
\beq
t=\sqrt{\frac53} \lambda \left| \frac{h_0}{a} \right |
\quad ; \quad
s = -\frac{2mh_0\lambda}{3a^2}t 
\quad ; \quad
u_k = -\frac{2mh_0\lambda}{ab_k}t  \  .
\label{stum}
\eeq
{}From the cubic equation we find the value $\lambda=(10)^{2/3}/20$.

As we advanced, with $m\not=0$  one can locate the minima in 
perturbative regions. Consider for instance the case $e_0=c_1=0$
and $\lambda_0=1$ so that the real moduli are given in (\ref{stum}).
Note that one can have $h_0$, $h_k$ and $c_2$ arbitrarily large
as long as $\lambda _0=1$ and  eq.(\ref{finetune}) is respected.
Then one can check that 
\beq
e^{\phi_4} \ \simeq \ h_0^{-2} 
\quad ; \quad  e^{\phi} \ \simeq \ h_0^{-1/2}   \  ,
\label{perturb}
\eeq
so that for sufficiently large $h_0$ the minima will be perturbative. 
Note also that the NS flux density
is diluted for large fluxes since it goes like $h_0/t^{3/2}
\simeq h_0^{-1/2}$. Concerning the RR flux $\ov{F}_2$, its density
also goes like  $h_0^{-1/2}$ for large fluxes, taking the factor of $g_s$
into account \cite{DeWolfe}. The cosmological constant
eq.(\ref{cosmo2}) scales like 
\beq
V_0\ \simeq -\ h_0^{-5} \ \simeq \  -(e^{\phi})^{10}
\eeq
and hence is substantially suppressed for large $h_0$.
Similar results may be obtained for values of $\lambda_0$
sufficiently close to 1, which would allow for contributions
to RR tadpoles with either sign and of arbitrary size.
In section \ref{sec:models}
we will consider this possibility to construct a 
semi-realistic intersecting D6-brane model with all
diagonal closed string moduli stabilized. 

Let us mention for completeness other solutions within this 
class of AdS minima with $m\not=0$. We may start by choosing
a preferred value for some of the moduli. For example, 
we can set $v=0$, and $h_0 \im S -\sum_k h_k \im U_k=0$.
Then, necessarily $c_1=-3h_0 c_2/a$ and, from the cubic  
equation with  $\lambda=-\lambda_0$, $e_0=45h_0 c_2^2/ma$.
This is the solution found in \cite{vz}.

Another way to proceed with the analysis is to fix $\lambda_0$.
For example, we can take $c_2=0$ so that $\lambda_0=0$. Obviously, 
$t^2 >0$ then requires either $\lambda<0$ or $\lambda > 1$. If $\lambda < 0$,
then $s >0$, $u_k>0$ demand $h_0m >0$ and $h_k m < 0$, thus the flux contribution to
tadpoles is like that of D6-branes. It is more interesting to consider
$\lambda > 1$ so that  $h_0m <0$ and $h_k m > 0$. Furthermore, to satisfy
(\ref{cubice}),
it must be  $(e_0 a - c_1 h_0) > 0$. Were it not for the fact that the fluxes are
integers, we could always find solutions for some chosen $\lambda$. But still
there is room to adjust the fluxes. For example, for $\lambda=3/2$ we just
need  $(e_0 a - c_1 h_0)= - 6 h_0^3 m/a^2$ and this could be verified say
for $e_0=0$, $c_1=24m$ and $h_0=2a$.

We can also set $h_0=h_k=0$, but to this end a different parametrization
of $v$, amounting to $\lambda \to \lambda_0 \hat{\lambda}$, must be 
employed. Now the interesting case is $\hat{\lambda} < -1$ because
$s >0$, $u_k>0$ require  $c_2 a>0$ and $c_2 b_k <0$  so that
the flux contribution to tadpoles could cancel that of D6-branes.
Again we can choose some $\hat{\lambda}$ and find values of $e_0$
to satisfy the cubic equation for $v$. For instance, for
$\hat{\lambda}=-3/2$ we need $e_0 m^2= - 161 c_2^3$.
One can check however that in this and the previous solution it is
again hard to achieve at the same time a large value
for the volume and a small value for the 10-dimensional dilaton,
the reason being that now the value of fluxes $h_0$, $c_2$ and $h_k$ will be
constrained by RR tadpole cancellation conditions.

One can also easily find 
non-susy AdS vacua. We will just show a particularly
simple example. In general, there are   
solutions in which (\ref{fixsu}) and (\ref{finetune})
are still satisfied. To go further let us set $m=0$. Then there are solutions
with $as=2 c_2 t$ and $v=h_0/3a$, but with the novelty that
\beq
\pm 9c_2 t^2 = e_0 - \frac{h_0 c_1}{a} -
\frac{h_0^2 c_2}{3a^2} \ .
\label{valtmnosusy}
\eeq 
With plus sign this is the supersymmetric minimum, but we can also choose the
minus sign depending on the fluxes. For instance, if $e_0=c_1=0$, only the
non-supersymmetric choice is available. In this case the minimum is AdS
and it is typically stable because the eigenvalues of the Hessian are positive
or negative but above the Breitenlohner-Freedman bound \cite{Breitenlohner}.

\section{D6-branes, fluxes and the Freed-Witten anomaly}
\label{sec:fw}

We are going to consider now adding D6-branes to a IIA
background with fluxes turned on.  Besides the general
RR tadpole cancellation constraints already mentioned, a number of
points do also change.  One apparent puzzle is the following.  In the
world-volume of a generic stack of D6-branes there is a $U(1)$ gauge
field whose scalar partner parametrizes the D6-brane position in
compact space. These $U(1)$'s often get St\"uckelberg masses by
swallowing RR scalars and disappear from the low-energy spectrum.  
At the same time these scalars participate in the cancellation of
$U(1)$ gauge anomalies through a variation of the Green-Schwarz mechanism
\cite{anomalous}. Let us specify now to the toroidal case considered in detail 
in the present paper. Consider a $\D6_a$ wrapping a factorizable torus with
wrapping numbers as in eq.~(\ref{wns}). Then the $U(1)_a$ field
couples to RR fields in \deq4 as follows
\beq F^a \ \wedge \ N_a\sum_{I=0}^3 \ c_I^a C_I^{(2)}
\label{bf}
\eeq
with
\beq c_0^a=m_a^1m_a^2m_a^3 \ ;\ c_1^a=m_a^1n_a^2n_a^3 \ ;\
c_2^a=n_a^1m_a^2n_a^3 \ ;\ c_3^a=n_a^1n_a^2m_a^3 \ \ .  
\eeq
Here the $C_I^{(2)}$ are 2-forms which are Poincar\'e duals in \deq4 to
the $\im U_I$ fields considered above. In terms of them the couplings
have a Higgs-like form
\beq A_\mu ^a \partial ^\mu 
( c_0^a\ \im S- c_1^a \im U_1 - c_2^a \im U_2 - c_3^a \im U_3) \ .
\label{higgs}
\eeq
We thus observe that certain linear combinations of $\im U_I$ ($U_0=S$) fields
get a mass by combining with open string vector bosons living on the
branes.  Moreover, these linear combinations of $\im U_I$ fields will
transform with a shift under $U(1)_a$ gauge transformations, like
Goldstone bosons do.  On the other hand, we have seen above that NS and metric
backgrounds give rise to terms in the superpotential linear in
$\im U_I$, i.e.
\beq 
W_Q\ =\ \int_{Y}\ \Omega_c \wedge (\ov{H}_3\ +dJ_c)\ = \
-\sum_{I,J=0}^3 A_{IJ} \wt{T}_I \wt{U}_J \ .  \eeq
Such terms generically may give rise to potential terms for the $\im
U_I$ fields which would not be invariant under the shifts induced by
$U(1)_a$ gauge transformations.  The condition to restore consistency
and gauge invariance would be to impose the constraint
\beq \int_{\Pi_a}\ (\ov{H}_3\ + \omega J_c) \ = \ 0 \  ,
\label{fwmetric}
\eeq
evaluated at the appropriate vacuum.  Here $\Pi_a$ denotes the 3-cycle
wrapped by the D6-brane, $\omega $ are the metric fluxes and
$J_c$ is the complexified K\"ahler 2-form of the torus. We have used
$dJ_c=\omega J_c$, according to (\ref{dXomega}).

In absence of metric fluxes eq.~(\ref{fwmetric}) may be understood in
terms of the Freed-Witten anomaly \cite{fw,mms,cu}\footnote{We thank
A. Uranga for pointing out this connection to us.}. A simple way to
see the origin of this constraint starting from type IIB is as follows
\cite{cu}. Consider a D3-brane wrapping a 3-cycle $\Pi$ through which
there is some quantized NS flux $\ov{H}_3$. On the world-volume of the
D3-brane there is a CS coupling of the form $\int C_2 \wedge F_2$,
$C_2$ being the RR 2-form and $F_2$ the open string gauge field
strength. After performing a IIB S-duality transformation one gets a
coupling
\beq 
\int _{\Pi\times {\bf R}} H_3 \wedge {\tilde A}_1 \  ,
\eeq
where ${\tilde A}_1$ is the gauge field dual to the $A_1$ form living
on the D3-brane. This shows that a background for $H_3$ gives rise to
a tadpole for ${\tilde A}_1$ and hence to an inconsistency.  Performing 
three T-dualities one expects for D6-branes the analogous term
\beq 
\int _{D6} H_3 \wedge {\tilde A}_4 \ .  
\eeq
The resulting tadpole is avoided if $\int_{\Pi _a} \ov{H}_3=0$. Equation
(\ref{fwmetric}) should thus be the extension of this constraint to
the case including metric fluxes $\omega \not= 0$.

Note that, ignoring for the moment the effect of metric fluxes,
eq.~(\ref{fwmetric}) implies that all D6-branes wrapping the
orientifold should obey
\beq 
\sum_{I=0}^3\ c^a_I \ h_I \ =\ 0 \ \ .
\label{fwdir}
\eeq
This is in general  a strong constraint on the possible D6-branes one
may add in specific models with flux, as we discuss below in specific
examples. Remarkably, this constraint guarantees that combinations of
axions getting masses by mixing with vector bosons are orthogonal to 
those becoming massive from fluxes, the latter being typically of the form
$h_0 \im S -\sum_k h_k \im U_k$.

Another interesting point is the following.  We have seen in
section~\ref{ssadsf} that adding fluxes one can
fix the torus moduli in a supersymmetry-preserving AdS minimum. Now, for non-zero
NS fluxes one finds at the minima that $h_i/h_0 = -
s/u_i$. Substituting this in eq.~(\ref{fwdir}) and multiplying by the
torus volume one arrives at
\beq 
m_a^1m_a^2m_a^3 (R_y^1R_y^2R_y^3)\ -\ m_a^1n_a^2n_a^3
(R_y^1R_x^2R_x^3)\ -\ n_a^1m_a^2n_a^3 (R_x^1R_y^2R_x^3)\ -\
n_a^1n_a^2m_a^3 (R_x^1R_x^2R_y^3) \ =\ 0 \ .
\label{slag}
\eeq
This condition means that the D6-brane wraps a special Lagrangian cycle (slag).  
From the effective Lagrangian point of view this 
is proportional to a Fayet-Iliopoulos 
(FI) term \cite{cim1} and hence it imposes dynamically that the D6-brane
configuration should be supersymmetric (i.e. all FI terms should
vanish). Thus one concludes that, in this class of AdS supersymmetric minima the
constraint (\ref{fwdir}) implies that the brane configuration should
be also supersymmetric.  
Notice that including metric fluxes in this class of minima does not add extra 
constraints to be satisfied due to the relations (\ref{finetune}).

The condition (\ref{slag}) in these  AdS vacua  is in fact
\beq
\im \Omega|_{\Pi_a} = 0  \  .
\label{slag2}
\eeq
It arises from  $\int_{\Pi _a} \ov{H}_3=0$ because at the AdS minimum
$\ov{H}_3 \propto \im \Omega$. In turn, in our setup this is a simple
consequence of $h_i/h_0=-s/u_i$. This is also found in a more general
analysis of type IIA susy AdS vacua \cite{bc, lt}.  
Likewise, if there are metric fluxes,
$dJ_c \propto \im \Omega$. For instance, in the models
of section \ref{ssadsf} this can be deduced using (\ref{fixsu}). Therefore, even if
the NS fluxes vanish, in these models there is still a FW constraint of 
the form
\beq
3a c_0^a - b_1 c_1^a - b_2 c_2^a - b_3 c_3^a = 0 \  . 
\label{fwb}
\eeq
This guarantees that combinations of axions acquiring a mass from fluxes
or from $U(1)$ mixing are orthogonal to each other

Recently models of type IIB orientifolds with fluxes and intersecting
(or rather magnetized) D-branes with semi-realistic spectrum have been
constructed \cite{blt, cu, lrs1, ms, Cveticflux}. Some of them do not verify the (IIB
version of) constraint (\ref{fwdir}) and hence would be in principle
inconsistent. This possible problem with the FW anomaly was already pointed out in
\cite{cu} where it was suggested that it could
be cured if additional D-branes were included. In the case of IIA
orientifolds under consideration we would need to add D4-branes
hanging between different sets of D6-branes and their orientifold
mirrors. It may be argued \cite{cu} that the chiral spectra from
intersecting D6-brane models does not get affected by the presence of
these extra D4-branes. However, no specific construction with this
possible cancellation mechanism has been presented in the
literature. In addition, it is not clear whether in the case of supersymmetric
D6-brane configurations the addition of the extra D-branes does not
spoil supersymmetry. Given this fact, it seems sensible to impose the
constraint (\ref{fwdir}) on specific models with Minkowski vacua and that will be 
our approach below. In AdS vacua, in which the real parts of all moduli are determined, 
the FW anomaly cancels automatically if the brane configuration is supersymmetric.

\section{Intersecting D6-brane models in the presence of fluxes}
\label{sec:models}

We have shown in previous sections that the addition of fluxes 
in type IIA theory leads to new properties not present in
analogous IIB models.
Some of the aspects we have found with potential
model-building applications are: 1) fluxes may contribute to all four  
RR tadpoles, 2) one can have examples of fluxes fixing 
part or all of closed string moduli but
not contributing to RR tadpoles, and 3)
there are  models with metric fluxes (as well
as other NS and RR fluxes) in which one can obtain AdS supersymmetric vacua    
with all moduli stabilized and contribution to RR tadpoles {\it       
opposite to that of D6-branes}.
In addition to  these properties, since plenty of flux variables do not 
contribute to RR tadpoles, there is substantial freedom in the choice of
the parameters of the  vacua and in particular one can obtain minima at large volume
and small dilaton values, in which the approximations inherent to a 4-dimensional
effective Lagrangian approach hold.

To illustrate the possible applications to model-building of these 
results in previous sections we are going to consider here specific  
 intersecting D6-branes models with semi-realistic
spectrum.  The first two examples correspond to  Minkowski vacua
both with unbroken \neq1 supersymmetry and with broken supersymmetry but no-scale
structure. Although in these cases only some of the closed string 
moduli are fixed at the minima, it is interesting to consider them
since other effects could perhaps stabilize the rest of the moduli.
In these two cases the models will be  
left-right symmetric extensions of the MSSM, with gauge
group $SU(3)\times SU(2)_L\times SU(2)_R\times U(1)_{B-L}\times U(1)$,
rather than the MSSM.
The third example has an  AdS supersymmetric background and is particularly interesting
since, to our knowledge, is the first semi-realistic three-generation model with 
all closed string moduli stabilized. In this case also the gauge group
is closer to that of the MSSM, since it is  that of
the SM with some additional $U(1)$'s. 
We  consider models with
non-supersymmetric intersecting branes and all closed string moduli fixed in
appendix B.

\subsection{Minkowski MSSM-like}
\label{ssminkmodels}

In the class of type IIB orientifold models with fluxes 
studied up to now, it has been shown that flux backgrounds with Minkowski
geometry, either \neq1 supersymmetric or not, lead to positive  
contributions to RR tadpoles. This stems from the fact that ISD fluxes
always contribute to RR tadpoles as D-branes do.
In building semi-realistic models this leads 
to problems with RR tadpole cancellation conditions, 
since typically fluxes contribute too much to tadpoles.
It was pointed out in  \cite{ms} that this problem may be 
cured if appropriate additional D9-anti-D9-brane pairs 
contributing negatively to some of the RR tadpoles are added. In any case, 
full cancellation of RR tadpoles in realistic toroidal models require
considering orbifold generalizations like $\Z_2\times \Z_2$ \cite{csu, Cveticz2z2}.
Semi-realistic \neq1 supersymmetric  
type IIB $\Z_2\times \Z_2$ orientifolds with flux backgrounds have 
been studied in \cite{blt, cu, lrs1, ms, Cveticflux}. The class of models of \cite{ms} has 
a brane content as given in table \ref{msmodels}.
\begin{table}[htb] \footnotesize
\renewcommand{\arraystretch}{1.25}
\begin{center}
\begin{tabular}{|c||c|c|c|}
\hline $N_i$ & $(n_i^1,m_i^1)$ & $(n_i^2,m_i^2)$ & $(n_i^3,m_i^3)$ \\
 \hline\hline $N_a=8$ & $(1,0)$ & $(3,1)$ & $(3 , -1)$ \\ $N_b=2$ &
 $(0,1)$ & $ (1,0)$ & $(0,-1)$ \\ $N_c=2$ & $(0,1)$ & $(0,-1)$ &
 $(1,0)$ \\ \hline \hline $N_{h_1}=2$ & $(-2,1)$ & $(-3,1)$ & $(-4,1)$
 \\ $N_{h_2}=2$ & $(-2,1)$ & $ (-4,1)$ & $(-3,1)$ \\ $8N_f$ & $(1,0)$ &
 $(1,0)$ & $(1,0)$ \\ \hline \end{tabular}
\end{center}
\caption{\small  Wrapping numbers giving rise to a MSSM-like spectrum. 
Branes $h_1$, $h_2$ and $o$ are added in order to cancel RR tadpoles.}
\label{msmodels}
\end{table}
In the case of the IIB $\Z_2\times \Z_2$ orientifold the 
$(n,m)$ integers would be magnetic numbers whereas in the
T-dual IIA orientifold they correspond to wrapping numbers 
along horizontal and vertical directions of each $\T^2$ in the factorized
$\T^6$ respectively.
Note that this set contains as a subset the MSSM-like model introduced
in \cite{cim2, cimyuks}. We assume as 
in \cite{cimyuks} that the $b$ and $c$ D6-branes 
sit on top of the orientifold plane so that the corresponding gauge  
symmetries are enhanced to $SU(2)_L$ and $SU(2)_R$ respectively. The
full initial gauge group is then $U(4)\times SU(2)_L\times SU(2)_R   
\times [U(1)_1\times U(1)_2]$.  Separating one of the $a$-branes from 
the other three produces the breaking $U(4) \rightarrow U(3)\times  
U(1)$.  Furthermore, two out of the three
$U(1)$'s get a St\"uckelberg mass by combining with RR axion fields.
We are thus left with a gauge group $SU(3)_c\times
SU(2)_L \times SU(2)_R\times U(1)_{B-L} \times [U(1)]$, which
contains the  left-right symmetric extension of the SM with  an extra  
$U(1)$. 
The branes $a,b,c$, give rise to a 3-generation MSSM-like spectrum  
whereas the  additional branes $h_{1,2}$ in table \ref{msmodels} 
are used to help in cancelling the RR tadpoles. 

Note that in
the case of the $\Z_2\times \Z_2$ IIA orientifold the RR tadpole cancellation 
conditions in the presence of fluxes will have the form
\beqa
\sum_a N_a n_a^1 n_a^2 n_a^3 + \oh(h_0 m + a_1 q_1 + a_2 q_2 + a_3 q_3)& = &
16 \ \ , \nonumber \\[0.2cm]
\sum_a N_a n_a^1 m_a^2 m_a^3 + \oh (m h_1 - q_1 b_{11} - q_2 b_{21} -
q_3 b_{31}) & = & -16\ , \nonumber \\[0.2cm]
\sum_a N_a m_a^1 n_a^2
m_a^3 + \oh( m h_2 - q_1 b_{12} - q_2 b_{22} - q_3 b_{32}) & = & -16  \ ,
\label{tadodhz2z2} \\[0.2cm]
\sum_a N_a m_a^1 m_a^2 n_a^3 + \oh (m h_3 - q_1 b_{13} - q_2 b_{23} -
q_3 b_{33}) & = & -16 \ . \nonumber
\eeqa
where the $(-16)$ in the last three conditions is the RR tadpole contribution of the 
other 3 orientifold planes existing in the $\Z_2\times \Z_2$ case.
Note that the branes $h_1$ and $h_2$ contribute negatively to all four RR tadpoles
so that in principle one can use them to compensate for a  too large 
contribution to the first tadpole condition from fluxes. Precisely this was
the approach in ref.\cite{ms} (see also \cite{Dudas}).
 Here we will use this class of models
as our starting point for the IIA orientifold case. Here are some
possibilities:

\noindent
{\it i) A 3 generation \neq1 MSSM-like model with some fixed moduli}

\noindent
Consider the above model in which we turn on non-vanishing fluxes 
as in one of the susy Minkowski examples of section 
(\ref{ssminkowski}) with non-vanishing $b_{31},b_{21},a_2,a_3$ (example {\it SM-2}).
The addition of NS fluxes $h_0,h_1$ and RR $e_0,e_2,e_3$, is optional, but we set
all the remaining backgrounds to zero. The superpotential has then the 
form
\beq
W = -T_2(a_2S+b_{21}U_1) - T_3(a_3S+b_{31}U_1)  + e_0 + ih_0S - 
ih_1U_1  +  i e_2T_2  + ie_3T_3  \  .
\label{supss1}
\eeq
As explained in section \ref{ssminkowski} this has a Minkowski 
supersymmetric minimum with
\beq
h_0=a_2 v_2 + a_3
v_3 \quad ; \quad e_2=a_2 \im S + b_{21} \im U_1 \quad ; \quad t_3=-\frac{b_{21}
t_2}{b_{31}} \quad ; \quad s=-\frac{b_{21} u_1}{a_2} \ .
\label{supss1cons}
\eeq
as long as $e_2a_3=e_3a_2$, $h_0b_{31}=-a_3h_1$, $h_0b_{21}=-a_2h_1$ and
$h_0e_2=e_0a_2$. Thus in this supersymmetric Minkowski background 
two complex linear combinations of moduli are fixed at the 
minimum.

Note that, since $m=q_i=0$, in this background the fluxes do not contribute 
to the RR tadpole. Thus one can consider the addition of 
D6-branes as in the the case
with $N_f=5$ in table \ref{msmodels}. As pointed out in \cite{ms}
with this choice all RR tadpoles cancel without the addition of fluxes 
in type IIB theory. In the present IIA case we can rather add the
background considered here and the RR tadpoles are not modified and
hence cancel. However the moduli are partially fixed 
by eq.(\ref{supss1cons}). 

It is easy to check that the $a,b$ and $c$ branes where the SM lives
trivially satisfy the FW constraint. However the branes
of type $h_{1,2}$ may be problematic unless:
\beq
 a_2(m_a^1m_a^2m_a^3) - b_{21}(m_a^1n_a^2n_a^3) \ = \ 
a_2 - 12 b_{21}\ =\ 0 
\eeq
which on the other hand may be easily satisfied
by appropriately choosing $a_2$, $b_{21}$. Note that this condition
guarantees
that the linear combination of $\im S$, $\im U_1$ getting masses through 
fluxes (eq.(\ref{supss1cons})) is orthogonal to the linear 
combination getting masses by mixing with the $U(1)$'s
of branes $h_{1,2}$. 

Note that in the IIB version of this orientifold with fluxes
considered in \cite{ms},
the latter contributes to RR tadpoles and one can only get
a one-generation \neq1 supersymmetric model. 

\noindent
{\it ii) A 3 generation no-scale  model}

\noindent
One can also consider one of the no-scale backgrounds discussed in section
(\ref{ssnoscale}), the variant of the {\it NS-2} model,  
and include a  set of D6-branes as in table 
\ref{msmodels}. A simple  example is as follows. 
Take non-vanishing $a_3,q_3$ with the remaining $q_i=a_i=0$.
In addition one may include non-vanishing $h_0,e_0,e_i$ but set the
remaining backgrounds to zero. 
The superpotential has then the form
\beq
W(S,T_i) = -a_3ST_3 - q_3T_1T_2  + e_0 + ih_0S
 + i\sum_ie_iT_i  \  .
\label{supss2}
\eeq
The imaginary part of $S$ and the 
$T_i$ are fixed as in eq.(\ref{min2c1}) whereas one has for the 
real parts the relationship $a_3st_3=q_3t_1t_2$. In addition one
has the constraint $e_0=h_0 \im S+e_1v_1$.
There is only a contribution equal to $\frac{1}{2}a_3q_3$ to the first
RR tadpole. We consider fluxes quantized in
units of 8 to avoid problems with
flux quantization \cite{blt, cu}. 
One can then cancel tadpoles in a $\Z_2\times \Z_2$
orientifold with branes as in table \ref{msmodels} with 
$N_f=1$ and $a_3=q_3=8$.

One can also consider a no-scale model with a non-vanishing 
IIA mass parameter $m$ and with no metric fluxes, as described at the 
beginning of subsection 5.2. One takes non-vanishing $m$ and $h_0$.
In addition one can have non-vanishing $e_i,q_j$, verifying
$\gamma_i=me_i+q_jq_k=0$ ($i\not=j\not=k$). Setting $h_0=m=8$ 
and $N_f=1$ one cancels all tadpoles. Note that this model,
which has no metric fluxes, is the IIA mirror of a similar 
no-scale model considered in \cite{ms}. 

One can check however that both these no-scale models
 as they stand have FW anomalies. The danger comes from the $h_{1,2}$ branes which 
have a non-vanishing product $m_a^1m_a^2m_a^3\not= 0$. One possibility which might cure this
problem is if, as suggested in \cite{ms},  the brane $h_1$ recombines with the mirror 
of $h_2$ into a single (non-factorizable) D6-brane $h_1+h_2'$. 
One can in fact claim that this is the generic situation for branes like these
which do intersect. In this case,
since $h_1$ and $h_2'$ have equal and opposite $m_a^1m_a^2m_a^3$, the FW 
would cancel on the recombined brane. On the other hand it is not clear whether
after the addition of fluxes a flat direction in the effective potential
exists corresponding to the recombination of those branes. 
In the $\neq1$ supersymmetric AdS model which we describe next no such problem appears.

\subsection{A \neq1 MSSM-like model with all closed string moduli stabilized in
AdS}
\label{ssadsmodels}

The previous  intersecting brane  models  were able to combine a semi-realistic
spectrum with a partial determination of some closed string moduli.
We now show that all such moduli may be stabilized in the case of AdS 
vacua, thus providing, to our knowledge, the first semi-realistic
string model with all closed string moduli stabilized at
weak coupling.  Note first that 
in the past it has been argued 
that it is impossible to construct semi-realistic \neq1 supersymmetric
intersecting D6-brane models wrapping the IIA orientifold $T^6/(\Omega
(-1)^{F_L} I_3)$. The reason for this was essentially the impossibility
to cancel the 4 RR tadpole conditions simultaneously while maintaining
supersymmetry.  To obtain \neq1 supersymmetric models extra orbifold twisting
(e.g. $\Z_2\times \Z_2$, as in previous examples) had to be added, giving rise to extra
orientifold planes which help in the cancellation of RR tadpoles \cite{csu}.  
We will show here that one can build \neq1 supersymmetric configurations
in the purely toroidal orientifold in which those RR tadpoles may be
cancelled by the addition of NS/RR and metric fluxes. The role played
by additional orientifold planes in orbifold (e.g. $\Z_2\times \Z_2$)
models is here played by the additional fluxes which contribute like
orientifold planes.  At the same time those fluxes stabilize all
closed string moduli in AdS space. Moreover the complex structure
moduli are fixed at values which render the D6-brane configuration
supersymmetric. Notice that in the \neq1 supersymmetric models
previously considered in the literature those moduli where not determined
by the dynamics.

Let us consider the set of D6-branes wrapping factorizable cycles in
the orientifold as in table \ref{modelito}.
\begin{table}[htb] \footnotesize
\renewcommand{\arraystretch}{1.25}
\begin{center}
\begin{tabular}{|c||c|c|c|}
\hline $N_i$ & $(n_i^1,m_i^1)$ & $(n_i^2,m_i^2)$ & $(n_i^3,m_i^3)$ \\
 \hline\hline $N_a=4$ & $(1,0)$ & $(3,1)$ & $(3 , -1)$ \\ $N_b=1$ &
 $(0,1)$ & $ (1,0)$ & $(0,-1)$ \\ $N_c=1$ & $(0,1)$ & $(0,-1)$ &
 $(1,0)$ \\ \hline \hline $N_{h_1}=3$ & $(2,1)$ & $(1,0)$ & $(2,-1)$
 \\ $N_{h_2}=3$ & $(2,1)$ & $ (2,-1)$ & $(1,0)$ \\ $N_o=4$ & $(1,0)$ &
 $(1,0)$ & $(1,0)$ \\ \hline \end{tabular}
\end{center} 
\caption{\small A MSSM-like model with tadpoles cancelled by fluxes.
Branes $h_1$, $h_2$ and $o$ are added in order to cancel RR tadpoles.}
\label{modelito}
\end{table}
Note that this set only differs from the previous examples in the 
form of the additional branes $h_1,h_2$. Another difference is
that in our IIA case we have a purely toroidal (no $\Z_2\times \Z_2$) orientifold 
without further twisting. The corresponding chiral spectrum at the intersections is 
given in table \ref{spec}.
\begin{table}[htb] \footnotesize
\renewcommand{\arraystretch}{1.25}
\begin{center}
\begin{tabular}{|c|c|c|c|c|c|}
\hline Intersection & Matter fields & Rep.  & $Q_{3B+L}$ & $Q_1$ &
 $Q_2$ \\ 
\hline\hline 
$a-b$ & $ F_L$ & $3(4,2_L)$ & 1 & 0 & 0 \\
\hline 
$a-c$ & $F_R$ & $3({\bar {4}},2_R)$ & -1 & 0 & 0 \\ 
\hline
 $b-c$ & $H$ & $(2_L,2_R)$ & 0 & 0 & 0 \\ 
\hline \hline  
$a-h_1$ & $T_1 $ & $(4,{\bar 3}_1)$ & 1 & -1 & 0 \\ 
\hline 
$a-h_1'$ & $T_1'$ & $5(4,3_1)$ & 1 & 1 & 0 \\ 
\hline 
$a-h_2$ & $T_2 $ & $5({\bar 4},3_2)$ & -1 & 0 & 1 \\ 
\hline 
$a-h_2'$ & $T_2'$ & $({\bar 4},{\bar 3}_2)$ & -1 & 0 & -1 \\ 
\hline 
$b-h_2$ & $D_2 $ & $2(2_L,{\bar 3}_2)$ & 0 & 0 & -1 \\ 
\hline 
$c-h_1$ & $D_1$ & $2(2_R,{\bar 3}_1)$ & 0 & -1 & 0 \\ 
\hline 
$h_1-h_2'$ & $X$ & $4({\bar 3}_1,{\bar 3}_2)$ & 0 & -1 & 1 \\ 
\hline 
\end{tabular}
\end{center} \caption{\small  Chiral spectrum of the MSSM-like model.}
\label{spec}
\end{table}
In the table a prime indicates the mirror brane.  
The gauge group after separating branes and after two
of $U(1)$'s get St\"uckelberg masses is 
$SU(3)\times SU(2)_L\times U(1)_R\times U(1)_{B-L}
\times [U(1)\times SU(3)^2]$. Note that, unlike the case of
the $\Z_2\times \Z_2$  models above, one can make the breaking
$SU(2)_R\rightarrow U(1)_R$ by brane splitting, and hence the 
gauge group is that of the MSSM supplemented by some extra $U(1)$'s. 
We have three generations of quarks and leptons, one Higgs multiplet $H$ 
and extra matter fields involving the auxiliary branes $h_1$, $h_2$ and
$o$.\footnote{One can check that if the branes $h_1$ and $h_2'$
recombine, most of the extra matter beyond the SM disappears from
the massless spectrum, with only additional $SU(2)_{L,R}$
doublets remaining.}

With this brane content (plus the mirrors) the RR tadpole cancellation
conditions are
\beqa 
64\ +\ \oh (h_0 m+ a_1 q_1 +a_2q_2 +a_3q_3)& = &16 \ , \nonumber  \\[0.2cm] 
-4 \ + \ \oh (h_1 m - q_1 b_{11} - q_2 b_{21} - q_3 b_{31})  & = & 0 
\ , \nonumber \\[0.2cm] 
-4 \ + \ \oh(h_2 m - q_1 b_{12} - q_2  b_{22} - q_3 b_{32}) & = & 0 \ ,
\label{tadodhhh} \\[0.2cm]
-4 \ + \ \oh (h_3 m - q_1 b_{13} - q_2 b_{23} - q_3 b_{33}) & = & 0 \ . \nonumber 
\eeqa
We see that to cancel tadpoles the sign of the flux contribution must
be opposite to that of D6-branes. We will now consider a AdS
background with metric fluxes and $m\not= 0$ discussed in section
(\ref{ssadsf}). 
The reader can check that choosing the fluxes as
\beq 
q_i=q=h_i-2 \quad ; \quad a_i=16 \quad ; \quad m=b_{ij}=-b_{ii}=4 \ ,
\label{canflux}
\eeq 
all RR tadpoles are cancelled. Note that eq.(\ref{finetune}) fixes $h_0=-12h_i$,
otherwise the values of $q$, $h_0$ and $h_i$ may be arbitrarily large still cancelling
all RR tadpoles. 

The above type of flux backgrounds does give rise to supersymmetric
AdS vacua with all real moduli fixed.  In
fact, the fluxes in (\ref{canflux}) are isotropic so that the
superpotential is of the form (\ref{sup2}) and the results of section
\ref{ssadsf} with $m\not=0$ can be applied. One can easily check that
with the above fluxes $\lambda_0=1+(24/h_0)$, which is arbitrarily close
to 1 for large $h_0$. Substituting these
fluxes yields for the real moduli
\beq 
s = -\frac{h_0\lambda}{96}t \quad ; \quad u_k = -\frac {h_0\lambda}{8} t
\quad ; \quad t = \sqrt{\frac53}\frac{|h_0|}{16}\lambda^{1/2}(\lambda+\frac 
{24}{h_0})^{1/2} \  ,
\label{gfixmod}
\eeq 
where $\lambda$ is the appropriate solution of eq.(\ref{cubice}) for 
the $\lambda_0$ indicated above.
For large $h_0$, $\lambda_0$ is close to 1 so that $\lambda \simeq (10)^{2/3}/20$
when $e_0=c_1=0$. In this case one needs $h_0<0$.
The imaginary part of the K\"ahler moduli are fixed as in eq.(\ref{newv})
whereas only the linear combination of dilaton and complex 
structure  axions $12 \im S +\sum_{k=1}^3  \im U_k$ is fixed, as in eq.(\ref{flatsu}).
As discussed in section \ref{ssadsf}, for large $h_0$ (which also implies
large $h_k$, $q$) all moduli are stabilized in a regime in which perturbation
theory in \deq4 is  a good approximation. 

Note that the FW conditions (\ref{fwdir}) 
for the D6-branes $a,h_1$ and $h_2$ read
respectively
\beq
h_2 \ =\ h_3   \quad ;\quad  h_1 \ =\ h_3  \quad ; \quad  h_1 \ =\ h_2  \  ,
\eeq
which are automatically satisfied because $h_1=h_2=h_3=
-h_0/12$. As we mentioned, this will guarantee that the 
the supersymmetry preserving conditions at the brane intersections
\beqa 
\atan \left(\frac{\tau_2}{3}\right)- \atan
\left(\frac{\tau_3}{3}\right)\ & = &\ 0  \  , \nonumber \\[0.2cm] 
\atan \left(\frac{\tau_1}{2}\right)- \atan\left(\frac{\tau_2}{2}\right)\ & = &\ 0  \  , 
\label{susyangles} \\[0.2cm] 
\atan \left(\frac{\tau_1}{2}\right)-
\atan \left(\frac{\tau_3}{2}\right)\ & = &\ 0   \  , \nonumber 
\eeqa
where $\tau_i=R_y^i/R_x^i$,
are satisfied, since  $u_1=u_2=u_3$.
This is no surprise, since as we mentioned in section \ref{sec:fw}, in this class
of AdS vacua all branes should be calibrated which in turn implies 
that the FW anomaly automatically cancels. 

In this particular model it is also interesting to look at 
the structure of $U(1)$'s and the $\im U_I$ RR fields.
It is easy to check that the couplings (\ref{bf}) give masses to
two linear combinations of $U(1)$'s by combining
with certain linear combinations of $\im U_I$ fields. Only the generator
$Q_a-2(Q_1-Q_2)$ remains massless at this level. 
On the other hand, the fields $\im S$ and $\sum_k \im U_k$ 
do not mix with the $U(1)$'s at all,
as expected, since FW anomalies cancel. Note that 
the combination $12\im S+\sum_k \im U_k$ is the one which gets 
a mass from fluxes (see eq.(\ref{flatsu})).
The orthogonal linear combination is massless and
may be identified with an axion which may be of relevance 
for the strong CP problem.

Although we have studied here only the dilaton 
and the diagonal closed string moduli of the orientifold,
we already mentioned that setting all off-diagonal 
moduli to zero solves the extremum conditions.
Furthermore, since we are in a \neq1 supersymmetric AdS background,
this guarantees that these off-diagonal moduli are also
stable. Thus, the closed string background discussed
is completely stable.
We have then succeeded in  building the
first  semi-realistic \neq1
supersymmetric model with all closed string moduli stabilized
in a consistent perturbative regime.
The vacuum is AdS with a c.c. which may be made small
(although not arbitrarily small, see below)
for large fluxes.
Unlike previous flux constructions in the
present case we have a simple toroidal orientifold, without any
further orbifold twist. Furthermore, the \neq1 supersymmetry
conditions on the brane angles are forced upon us by the Freed-Witten
constraint plus the minimization.  Recall in this respect that in the
\neq1 supersymmetric brane configurations constructed up to now the
angles were fine-tuned to verify the supersymmetry conditions, there was no
dynamical explanation for that choice, since not all closed string
moduli were fixed.

Let us make some complementary comments about this kind of MSSM-like AdS
constructions:

{\it i)} Other MSSM-like models in AdS may be constructed along similar lines 
making use of the backgrounds with metric fluxes and $m\not=0$ discussed
in subsection \ref{ssadsf}. An easy way to proceed 
is to start with a tadpole free \neq1 MSSM-like D6-brane
configuration and embed it in a AdS background of the type
$\lambda_0=1$  in which fluxes do not contribute to RR tadpoles 
at all. For example, one can start again from the $\Z_2 \times \Z_2$
orientifold example in table \ref{msmodels} with $N_f=5$ in which
all RR tadpoles cancel. In the prescribed AdS background all
moduli are stabilized in a perturbative regime for large enough 
fluxes.

{\it ii)}
One could think of building analogous AdS  MSSM-like models
with a background free of metric fluxes, as in
subsection \ref{ssadsnf}. This turns out to be difficult because
fluxes contribute negatively to the RR tadpoles in eq.(\ref{tadodhiso}) and 
hence additional orientifold planes have to be added wrapping those directions. 
One possibility is to use again the $\Z_2 \times \Z_2$ orientifold which has such 
O6-planes, but the fluxes tend to overwhelm the contribution of orientifolds
and this procedure does not look promising.

{\it iii)}
Besides the chiral spectrum described above, this class of toroidal models 
has massless adjoint chiral fields corresponding
to the open string moduli parametrizing the location and Wilson lines on
the branes. In a supersymmetric AdS background as the one we are considering here,
those open string moduli are in any case stable. It is an interesting question
to study what happens to them when some supersymmetry-breaking effect
is included.
It has been shown that fluxes in type IIB stabilize some
(but not all) of the open string moduli in the toroidal case \cite{openmoduli}. 
Additional ways to give masses to these degrees of freedom in toroidal models have been 
recently described in \cite{bcms, Angelantonj}. It should be worth to study this question 
in the context of our type IIA AdS backgrounds.

{\it iv) }
Once all moduli are fixed in a given model like this, one can compute 
a number of interesting physical quantities like gauge coupling 
constants and Yukawa couplings, since they will be known functions of
the fluxes. For example, in the above model the gauge kinetic 
functions of the  groups $SU(4)$, $SU(2)_L$, $SU(2)_R$ have
\beq
\re f_{SU(4)} \ =\ 9s+u_1 \quad  ;\quad  
\re f_{SU(2)_L}\ =\ \frac {u_2}{2} \quad  ;\quad 
\re f_{SU(2)_R}\ =\ \frac {u_3}{2}  \ .
\label{acoplosguay}
\eeq
Since these are the values at the string scale, to make 
contact with experiment we should then consider the running to
low energies. As we said, simplest toroidal models like this
have, in addition to the chiral spectrum, adjoint chiral fields
which will generically spoil the running of coupling constants.
Let us nevertheless proceed and compute them in this example.
Since $u_k$ and $s$ are related by (\ref{fixsu}) one has e.g.
\beq
\alpha_{SU(2)_L}\ =\ \frac {8}{\lambda |h_0|t} \ =\ 
\frac {128 \sqrt3}{\sqrt{5} h_0^2 \lambda^{3/2}(\lambda+24/h_0)^{1/2}}\  , 
\label{alpha2}
\eeq
and $\alpha_{SU(4)}=\frac27 \alpha_{SU(2)_L}$. Thus, 
we see that the  SM gauge coupling constants depend strongly on the fluxes.
In particular  we cannot make $h_0$ arbitrarily large (as one would naively 
do to decrease the value of the c.c.) since we would get then too
small SM gauge couplings, inconsistent with experiment.
For example, in the present model 
one can see that in order to get values $\alpha_i\simeq 1/5-1/30$ which
might be consistent with low energy physics one needs to have $h_0\simeq 100$
but not much bigger. This seems to be a generic property  and not a
particular feature of this class of models. Thus,
indeed we have an infinite  `landscape' of models depending on unconstrained
fluxes, but only a narrow region of fluxes would lead to 
consistent low-energy physics.
Something similar happens with the Yukawa couplings, which have been computed and 
can be neatly written in terms of products of Jacobi $\vartheta$-functions 
in this model \cite{cimyuks, yuks}. 
They scale like the gauge couplings and hence are equally suppressed 
for large fluxes.

{\it v)}
These  models are constructed in AdS and an obvious question is how one
could promote this kind of vacua to dS. One possibility which comes to
mind is to add  anti-D6-branes.
Indeed, if we add a pair of ${\rm D6-}{\overline {\rm D6}}$ branes to the
model,  there is an extra contribution to the scalar potential  which
has the form 
\beq
V_{\overline {D6}}\ \sim \ \frac {1}{u_1u_2u_3}
\eeq
and should be included in the complete minimization.
This would be essentially the mirror of the  approach in \cite{kklt} 
which was used in the type IIB case.
More generally, one may consider sets of D6-branes with
uncancelled NS-tadpoles. A potential is generated due to the 
missancellation of the tensions of the D6-branes against the 
orientifold tension, i.e. (in the string frame)
\beq
V_{\rm D6/O6} \ = \ {T_6 \over  {g_s}} \left( \sum_a N_a \A l_a \A -
\A l_{ori} \A\right) \ > \ 0
\label{NStad}
\eeq
where $\A l_a \A$ $(\A l_{ori} \A)$ are the volume of the 3-cycles wrapped
by each D6-brane (orientifold). It remains to be seen whether such
a procedure can be made to work. In any event, the kind of
fluxes considered here will stabilize non-supersymmetric D6-brane configurations
with non-vanishing NS-tadpoles that have been considered in recent
years. Examples of such non-supersymmetric D6-brane configurations with the
chiral content of the SM are presented in Appendix B.

\section{Conclusions}

In this paper we have studied the minima of the flux-induced 
effective moduli potential in a simple $\T^6/(\Omega (-1)^{F_L} I_3)$ 
IIA orientifold. 
We have focused on the dilaton and the diagonal K\"ahler and 
complex structure fiels, but we have nevertheless argued that
the results found ignoring off-diagonal moduli still provide stable extrema 
of the full potential in relevant cases.
We have considered RR, NS as well as metric background 
fluxes. Unlike the IIB case, the richness of the flux options leads 
to a full stabilization  of all closed string moduli in AdS without the
need of non-perturbative effects. Furthermore, the RR tadpole conditions,
which are very restrictive in the IIB case, only constraint some flux
combinations  in the IIA case. Thus, there is enough freedom 
to adjust fluxes so that the minima are located in regions with
large volume and small dilaton where the effective 4-dimensional 
supergravity approximations hold.  The combination of metric fluxes
with NS/RR fluxes leads to new possibilities such as fluxes fixing all
moduli in \neq1 supersymmetric AdS but not contributing to RR tadpoles. 
This provides us with a rigid `corset-like' background which can stabilize any
RR tadpole-free D6-brane configuration in this toroidal setting. 
In general, if metric fluxes are turned on, the overall fluxes can contribute
to RR tadpoles like O6-planes do, thereby providing the interesting
possibility of disposing of orientifold planes in some cases.

In models with all real moduli fixed, only one linear combination of 
the axions of the dilaton and complex structure fields 
is determined at the minima. This, which at first sight appears to be a limitation of the
approach, is in fact a blessing. Indeed, eventually we may  like to add 
systems of  D6-branes leading to chiral physics in the
background. The RR axions which are never fixed by fluxes are in fact needed 
by the  D6-branes to get rid of (potentially anomalous) open string
$U(1)$'s. We have seen that cancellation of the Freed-Witten anomaly 
guarantees that sufficient axions remain to give St\"uckelberg masses to the
$U(1)$'s. In the case of AdS \neq1 supersymmetric AdS vacua the cancellation of 
FW anomaly does in turn force the different sets of D6-branes
to be calibrated.

One can construct explicit models with a chiral spectrum quite close to
that of the MSSM with three generations and with all closed string moduli 
fixed in AdS.  In its construction we make use of fluxes (including metric ones)
contributing like orientifold planes to RR tadpoles. Other analogous models may also
be built. The minima may be located at large volume and small dilaton so that
we can trust our approximations. In such a model, with all moduli fixed,
one can compute explicitly all gauge and Yukawa couplings
as known functions of the fluxes undetermined by RR tadpoles. In the 
particular example of section \ref{ssadsmodels}, essentially only 
one flux (which may be identified with the NS 3-form flux $h_0$) 
fixes all couplings and scales. 
Thus, although one may talk about a landscape of models depending on
a single flux parameter $h_0$, only a narrow region of integer values
for $h_0$ would give rise to gauge couplings compatible with experimental
constraints. In particular, in our concrete model there is not enough freedom
to make the c.c. arbitrarily small by making $h_0$ large. We believe that this
is quite a generic feature. The dilaton and complex structure fields 
which determine the (inverse of) gauge coupling constants grow like 
some power of the fluxes. If we make the fluxes too large in order to get
e.g. a small c.c. the SM couplings would get far too small.

Our approach has been to consider the metric fluxes as a deformation added to
the original torus. The resulting twisted torus is a non-Calabi-Yau manifold
in which we still know the moduli and are able to introduce D6-branes. 
It would be interesting to go beyond the toroidal geometry in this spirit. 
In appendix A we have shown that the analysis of \neq1 vacua
deduced from the effective flux-induced superpotential agrees with recent
results on supersymmetric IIA compactifications on manifolds with
$SU(3)$ structure \cite{Gauntlett, Prezas, gmpt, Kaste, bc, lt, House}. 
Although we have worked out many specific minima of the general fluxed
potential in the simplest IIA toroidal orientifold, we cannot claim that
we have done a complete analysis. Furthermore, we have not explored
the possibilities offered by some of the solutions (e.g. 
AdS non-supersymmetric minima) that we have analyzed nor made a systematic
search for MSSM-like D6-brane configurations. Presumably there are many other
options beyond the ones that were discussed. All the models with all
moduli stabilized are however AdS. An important problem is how  to
modify the premises in order to obtain models with dS vacua. A possible option is to
add anti-D6-branes or, more generally, consider non-supersymmetric
brane configurations. Positive definite contributions to the potential will then 
in general appear which might help in going to dS. 
We hope to come back to all these issues  in the near future.

\vspace*{1cm}

{\bf \large Acknowledgments}

We thank  J.F.Garc\'{\i}a-Cascales, F. Marchesano,  
S. Theisen, and especially A. Uranga for useful discussions.
A.F. thanks the Max-Planck-Institut f\"ur Gravitationsphysik
for hospitality while preparing this paper. The work of 
P.G.C. is supported by  the  Ministerio de Educaci\'on y Ciencia (Spain)
through a FPU grant. 
This work has been partially supported by the European Commission under
the RTN European Program MRTN-CT-2004-503369 and the CICYT (Spain).

\newpage

\section*{Appendix A: $SU(3)$ structure of twisted torus}
\label{appA}
\setcounter{equation}{0}
\renewcommand{\theequation}{A.\arabic{equation}}

In this appendix we study the relation between the metric fluxes and the
$SU(3)$ structure of the twisted torus. The idea is to generalize
the analysis of \cite{glmw, kstt} by turning on all metric fluxes in (\ref{abmatrix})
and not only the $a_i$ obtained from T-duality of NS fluxes. We will
see that in this more general situation the twisted torus is still a half-flat
manifold as it occurs when only the $a_i$ are present \cite{glmw, kstt}.
Using the results in this appendix we will also be able to describe 
our Minkowski and AdS supersymmetric vacua in terms of torsion classes.
Supersymmetric IIA compactifications on manifolds with $SU(3)$ structure 
have been recently considered in 
\cite{Gauntlett, Prezas, gmpt, Kaste, bc, lt, House}.  

On the twisted torus one can build the fundamental 2-form $J$ and the
holomorphic 3-form $\Omega$ in the usual way. Including the sizes and 
complex structure parameters we have
\beqa 
J & = & -t_1 \eta^1 \wedge \eta^4 -t_2 \eta^2 \wedge \eta^5 
-t_3 \eta^3 \wedge \eta^6 \ , \nonumber \\[0.2cm] 
\Omega & = & 
(\eta^1 + i\tau_1\eta^4) \wedge (\eta^2 + i\tau_2\eta^5) 
\wedge (\eta^3 + i\tau_3\eta^6) \   .
\label{su3forms}
\eeqa 
These forms define an $SU(3)$ structure. In particular, they satisfy
\beq
J\wedge \Omega=0 \quad ; \quad  
J\wedge J \wedge J = -\frac{3i}{4} \frac{t_1 t_2 t_3}{\tau_1 \tau_2 \tau_3} \Omega \wedge \Omega^* \  .
\label{su3s}
\eeq  
The torsion classes can be read from (see e.g. \cite{glmw, ccdlmz, Gauntlett}) 
\beqa 
dJ & = & \frac32  \frac{t_1 t_2 t_3}{\tau_1 \tau_2 \tau_3} 
\im (\cw_1 \Omega^*) + \cw_4 \wedge J + \cw_3 \ , \nonumber \\[0.2cm] 
d\Omega & = & \cw_1 J \wedge J + \cw_2 \wedge J + \cw_5^*
\wedge \Omega \ ,
\label{djo}
\eeqa 
where $\cw_1$ is a complex 0-form, $\cw_2$ is a primitive
($\cw_2 \wedge J \wedge J =0$) complex 2-form, $\cw_3$ is a primitive
($\cw_3 \wedge J=0$) real $(2,1)\oplus(1,2)$-form, $\cw_4$ is a real
1-form, and $\cw_5$ is a complex $(1,0)$-form. The unusual factor
in the first term of $dJ$ is needed so that $d(J\wedge \Omega)=0$.

When only the metric fluxes (\ref{abmatrix}) are turned on, 
using (\ref{ocon}) we find 
\beqa 
d\Omega \!\!& = & \!\! \frac1{s}(a_1s + b_{11}u_1 + b_{12}u_2 + b_{13} u_3) \eta^{2536} + 
\frac1{s}(a_2s + b_{21}u_1 + b_{22}u_2 + b_{23}u_3) \eta^{1436}
\nonumber \\[0.2cm]
& {} & \ \ \ \ \ \  
+\frac1{s}(a_3s + b_{31}u_1 + b_{32}u_2 + b_{33}u_3) \eta^{1425} 
\ , \nonumber \\[0.2cm] 
dJ \!\! & = & \!\! (a_1t_1 + a_2t_2 + a_3t_3)\eta^{456}
- (b_{13} t_1 + b_{23} t_2 + b_{33} t_3) \eta^{126}
\label{djott}  \\[0.2cm]
& {} & \ \ \ \ \ \  
- (b_{12}t_1 + b_{22}t_2 + b_{32} t_3) \eta^{153} 
- (b_{11}t_1 + b_{21}t_2 + b_{31}t_3) \eta^{423}  
\ , \nonumber
\eeqa 
where $\eta^{153}= \eta^1 \wedge \eta^5 \wedge \eta^3$,
etc.. In $d\Omega$ we have used $u_i = s \tau_j \tau_k$, 
$i \not= j \not= k$.

Clearly, $d(J\wedge J)=0$ and $d(\im \Omega)=0$, thus the twisted
torus with the given fluxes is a half-flat manifold. For the torsion classes 
we easily read $\cw_4=\cw_5=0$. The torsions $\cw_2$ and $\cw_3$ are 
different from zero, they can be readily obtained using (\ref{djo}) and
\beqa 
\cw_1 & = & \frac1{6s t_1 t_2 t_3}
(a_1s t_1 + a_2 s t_2 + a_3 s t_3 + b_{11} t_1 u_1 + b_{12} t_1 u_2  + b_{13}t_1 u_3 
\nonumber \\[0.2cm] 
& {} & \ \  \ \
+ b_{21} t_2 u_1 + b_{22} t_2 u_2 + b_{23} t_2 u_3 
+ b_{31}t_3 u_1 + b_{32} t_3 u_2 + b_{33} t_3 u_3) 
\ . \label{torw1} 
\eeqa
Notice that this $\cw_1$ is similar to $W_Q$ without the NS fluxes, 
as expected because it is basically computed as $\int \Omega \wedge dJ$. 

In \cite{gmpt} it has been shown that supersymmetric Minkowski vacua
of type IIA require $\cw_1=0$. In our examples of this kind of vacua
in section \ref{ssminkowski} we indeed find $\cw_1=0$. 
This follows simply because taking real part of $\partial W/\partial \wt{U_J}=0$
gives $\sum_i A_{iJ} t_i =0$ which is enough to show $\cw_1=0$. In
fact, $dJ=0$ so that $\cw_3=0$ as well. We also find that $m=0$ and then
from the real part of $\partial W/\partial T_i=0$ we deduce that
$d\Omega=-\frac1{s} \ov{F}_2 \wedge J$. We have examples, such as
$W(S,T_1, T_2, T_3)$, $W(U_1,T_1, T_2, T_3)$ or the $W(T_1, T_2, T_3, U_2, U_3)$
in (\ref{w5f}), in which $d\Omega \not=0$ and $\cw_2=-\frac1{s}\ov{F}_2$.
Another characteristic feature of these models with $\ov{F}_2 \not=0$ is
the existence of flux tadpoles for some component of $C_7$. 
There are other models, such as  $W(S,U_1, T_2, T_3)$ or $W(U_1,U_2, T_1, T_2)$,
in which  $d\Omega=0$ and $C_7$ tadpoles vanish because
$\ov{F}_2 =0$ is required to have non-zero real parts of the moduli.
In all cases, the typical configuration has neither NS fluxes nor
RR fluxes for $F_6$ ($e_0=0$) and $F_4$ ($e_i=0$). These results are
in agreement with the analysis of \cite{gmpt, Kaste}. 

Type IIA supersymmetric AdS compactifications have also been studied
in terms of $SU(3)$ structures \cite{bc, lt}. It is interesting to
see how the same type of results follows in our setup. To begin
we notice that taking real part of $D_{\wt{U}_J}W=0$ gives
\beq
\sum_{i=1}^3 A_{iJ} t_i \wt{u}_J = -\oh \re W  \quad ; \quad J=0, \cdots, 3  \ .
\label{realDUW}
\eeq
We can use these relations to compute $dJ$ and also
\beq
\cw_1 =-\frac{\re W}{3 s t_1 t_2 t_3}  \  .
\label{w1ads}
\eeq
From the explicit $dJ$ we further read $\cw_3=0$. To calculate
$d\Omega$ we look instead at the real part of $D_{T_i}W=0$ and deduce
\beq
a_is + b_{i1}u_1+ b_{i2}u_2 + b_{i3}u_3 = -\frac{\re W}{2 t_i} - \sum_{j\not=k\not=i}
t_j(q_k + m v_k)  \  .
\label{realDTW}
\eeq
Moreover, combining with (\ref{realDUW}) yields the relation
\beq
t_1 t_2(q_3 + mv_3) + t_1 t_3(q_2 + mv_2) +t_2 t_3(q_1 + mv_1) = \frac14 \re W  \ .
\label{contqv}
\eeq
It is then straightforward to determine $d\Omega$ and from it obtain
\beq
\cw_2 \wedge J = -\frac14 \cw_1 J \wedge J + \frac{m}{s} B_2 \wedge J - 
\frac1{s} \ov{F}_2 \wedge J  \  ,
\label{w2ads}
\eeq
which satisfies $\cw_2 \wedge J \wedge J =0$ by virtue of (\ref{contqv}).
We conclude that supersymmetric AdS compactifications have $\cw_1$ and
$\cw_2$  different from zero but $\cw_3=\cw_4=\cw_5=0$, as found in
\cite{lt} in a more general setup. There is also a particular case
in which $\cw_2=0$ \cite{bc, lt}. Our example in section \ref{ssadsf}
is of this type. Indeed, using (\ref{fixsu}) gives $\cw_1=2a/t^2$ and
$d\Omega=\cw_1 J\wedge J$. We also find $\ov{H}_3 = -\sqrt{|h_1 h_2 h_3/h_0|} \im \Omega$.

\section*{Appendix B: Stabilizing non-susy intersecting D-brane models}
\label{ssfwmodels}
\setcounter{equation}{0}
\renewcommand{\theequation}{B.\arabic{equation}}

In the $\neq1$ supersymmetric AdS constructions in the main text,
we have discussed  examples in which
all D6-branes preserve the same \neq1 supersymmetry. In this appendix we would
like to study  the non-supersymmetric class of
semi-realistic intersecting D6-brane models of ref.~\cite{imr}.
One of the known problems of these non-susy models is that
they are unstable due to the
existence of NS tadpoles. These appear from a miscancellation of the
tensions of the D6-branes with the orientifold tension. We would like
to point out here that those non-susy models may in general become 
stable in  the presence of fluxes.
We will see that the FW conditions in AdS will force the branes to
preserve supersymmetry locally, i.e. any pair of intersecting D6-branes
will preserve one unbroken supersymmetry although there is no overall
\neq1 supersymmetry preserved simultaneously by all D6-branes.
We will see that all closed string moduli will be also determined,
although, as we argue at the end, a complete treatment would require
taking into account D-term's in the scalar potential.

In  \cite{imr} a general class of solutions was given for the wrapping
numbers $(n_a^i,m_a^i)$ giving rise to a SM spectrum. These are shown
in table~\ref{solution}. In this table we have several discrete
parameters.  First we consider $\beta^i =1,1/2$.  From the point of
view of branes at angles $\beta^i = 1$ stands for a rectangular
lattice for the $i^{th}$ torus, whereas $\beta^i = 1/2$ describes a
tilted lattice allowed by the $\Om I_3$ symmetry.  We also have two
phases $\epsilon, \tilde \eps = \pm 1$ and the parameter $\rho$ which
can only take the values $\rho = 1, 1/3$.  Furthermore, each of these
families of D6-brane configurations depend on four integers
($n_a^2,n_b^1,n_c^1$ and $n_d^2$).  Any of these choices leads exactly
to the same massless fermion spectrum of the SM with 3 generations.

\begin{table}[htb] \footnotesize
\renewcommand{\arraystretch}{2}
\begin{center}
\begin{tabular}{|c||c|c|c|}
\hline $N_i$ & $(n_i^1,m_i^1)$ & $(n_i^2,m_i^2)$ & $(n_i^3,m_i^3)$ \\
 \hline\hline $N_a=3$ & $(1/\beta ^1,0)$ & $(n_a^2, \epsilon \beta^2)$
 & $(1/\rho, - \tilde \eps/2)$ \\ \hline $N_b=2$ & $(n_b^1,\tilde
 \eps\epsilon \beta^1)$ & $ (1/ \beta^2,0)$ & $(1,-3\rho \tilde\eps
 /2)$ \\ \hline $N_c=1$ & $(n_c^1,3\rho \epsilon \beta^1)$ &
 $(1/\beta^2,0)$ & $(0,1)$ \\ \hline $N_d=1$ & $(1/\beta^1,0)$ & 
 $(n_d^2,\epsilon\beta^2/\rho)$ & $(1, 3\rho \tilde\eps /2)$ \\ \hline
 \end{tabular}
\end{center} \caption{\small D6-brane wrapping numbers giving rise to a SM
spectrum.  The general solutions are parametrized by a phase
$\epsilon, \tilde \eps =\pm1$, the NS background on the first two tori
$\beta^i=1-b^i=1,1/2$, four integers $n_a^2,n_b^1,n_c^1,n_d^2$ and a 
parameter $\rho=1,1/3$.}
\label{solution}
\end{table}
Now, imposing the condition (\ref{fwdir}) for branes $a$, $b$, $c$,
$d$, respectively, leads to the relations
\beqa
h_2 \ \frac {\epsilon \beta^2}{\rho \beta^1} \ -\ h_3 \ \frac
{{\tilde {\epsilon}}n_a^2}{2\beta^1}\ & = &\ 0 \ , \nonumber \\[0.2cm]
h_1 \ \frac {\epsilon {\tilde {\epsilon}}\beta^1}{ \beta^2} \ -\
h_3 \ \frac{{\tilde {\epsilon}}3\rho n_b^1}{2\beta^2}\ & = &\ 0 \ ,
\label{ligadosimr} \\[0.2cm]
h_3 \ \frac {n_c^1}{\beta^2}\ &=&\ 0   \ , \nonumber \\[0.2cm]
h_2 \ \frac {\epsilon \beta^2}{\rho \beta^1} \ +\ h_3 \
\frac {{\tilde {\epsilon}}3\rho
n_d^2}{2\beta^1}\ & = &\ 0  \ . \nonumber
\eeqa
These constraints have two solutions, depending on the value of
$n_c^1$:

\noindent
{\it i) $n_c^1\not=0$}

\noindent
In this case necessarily $h_3=0$ and hence $h_1=h_2=0$.  Only the flux
$h_0$ may be added and only the field $S$  may be fixed by
fluxes.

\noindent
{\it ii) $n_c^1=0$}

\noindent
In this case one can check that the constraints are solved as long as
\beqa
h_3 \ (n_a^2&+&3\rho n_d^2) \ = 0 \ , \nonumber \\[0.2cm]
h_1 \ & =& \ \frac {3\rho  n_b^1}{2\epsilon \beta^1} \ h_3 \ ,
\label{hicons} \\[0.2cm]
h_2 \ & =& \ \frac {\rho {\tilde {\epsilon}} n_a^2}{2\epsilon \beta^2} \ h_3 \ ,
\nonumber
\eeqa
so that one can only have non-vanishing $h_i$ if $n_a^2 = -3\rho
n_d^2$. One can check that, when this condition is verified, there are
two massless $U(1)$'s in the spectrum, $U(1)_R$ and $U(1)_{B-L}$,  
rather than just hypercharge, and only two linear combinations of the RR
fields
\beq
\im U_2 \ -\ \frac {{\tilde {\epsilon}} \rho n_a^2}{2 \epsilon
\beta^2} \im U_3 \quad ;\quad \im U_1 \ -\ \frac {3 \rho n_b^1}{2
\epsilon \beta^1} \im U_3
\eeq
become massive by combining with the $U(1)_{3B+L}$ and $U(1)_b$ gauge bosons
respectively.
The orthogonal linear combination
\beq 
\frac {3 \rho n_b^1}{2 \epsilon \beta^1} \im U_1 \ +\ \frac
{{\tilde {\epsilon}} \rho n_a^2}{2 \epsilon \beta^2} \im U_2 \ +\ \im
U_3 \ =\ \frac {1}{h_3} (\sum_{I=1,2,3} h_I \im U_I)
\eeq
is precisely a piece of the combination appearing in the superpotential, which is
expected to acquire a mass from fluxes upon minimization.  The other RR
field $\im S$ may become massive depending on the presence or not of
a non-vanishing $h_0$ background, which leads to no constraint in the 
model (since $m_a^1m_a^2m_a^3=0$ for all the D6-branes present).

If upon minimization one finds $\langle \re U_I \rangle \simeq 1/h_I$
 one obtains that the above conditions imply
\beq
\frac {\re U_3}{\re U_1} \ =\ \frac {3\rho
n_b^1}{2\epsilon\beta^1} \quad ;\quad \frac {\re U_3}{\re U_2} \ =\ \frac
{\rho {\tilde {\epsilon}} n_a^2}{2\epsilon\beta^2}  \  .
\eeq 
One can check that these conditions guarantee that at each brane
intersection there is one unbroken supersymmetry, although in this model no
overall supersymmetry generator is preserved by all intersections.
This kind of local (but not global) supersymmetry was termed `Q-SUSY' in \cite{cim1}.
Thus we
see that adding fluxes $h_i\not=0$, $i=1,2,3$, is only possible if the
brane configuration is locally supersymmetric.

Let us now be more explicit and chose the wrapping number parameters as follows:
\beq  
n_a^2=-n_d^2=\beta_1=\epsilon={\tilde {\epsilon}}=1 \quad  ;\quad 
\beta_2=1/2 \quad ; \quad \rho =1/3 \quad ; \quad n_b^1=2 \   .
\label{qsusycon}
\eeq
One can easily check that RR tadpoles cancel without the addition of any further
D6-brane nor fluxes. The conditions (\ref{hicons}) now read
\beq
h_1 \ =\ h_3\ =\ 3\ h_2  \  .
\eeq
Consider now again the same AdS vacua with $m\not= 0$
that  we discussed in section \ref{ssadsf}. We saw there that for
$m\not=0$ one can find AdS vacua in which the NS/RR contribution $mh_0$ to RR
tadpoles may be cancelled by the metric fluxes contribution $3ac_2$
($\lambda_0=1$ case).
Then we have the interesting possibility of fixing all closed string moduli
without fluxes contributing to RR tadpoles at all.
Consider backgrounds as follows
\beq
b_k\ =\ (-6,-2,-6) \quad  ; \quad  h_k \ =\ r\ b_k  \  ,
\eeq
where $r=-h_0/3a$ must be a positive integer. Then, if we further chose
$e_0=c_1=0$ one finds a minimum of the flux-induced potential
as long as $h_0>0$, $a<0$, $m<0$. The flux contribution to tadpoles
vanishes if we further take $c_2=-rm$. Then the real parts of
closed string moduli are fixed as 
\beq
u_k\ =\ \frac {3as}{b_k} \quad ; \quad  s\ =\ \frac {m}{a10^{1/3}} rt 
\quad  ; \quad  t\ =\ \frac {\sqrt{15} 10^{2/3}}{20} r  \  .
\eeq
Certain linear combinations of the imaginary parts of the
moduli fields are fixed as discussed in subsection \ref{ssadsf}.
Note that choosing $r=-h_0/3a$ large one can fix the moduli
at arbitrarily large values with small 4- and 10-dimensional
dilatons.

As we mentioned, one has to be careful in applying the results obtained in
the main text  to a non-supersymmetric brane configuration like this. Indeed,
in this case in addition to the F-term scalar potential one has to add the piece  
(\ref{NStad}). Still one expects a full determination of all closed string moduli
also in this non-supersymmetric example.

\end{document}